\documentclass[aps,twocolumn,nofootinbib,superscriptaddress,floatfix]{revtex4}
\usepackage{color,amsmath,amssymb,graphicx,latexsym,subfigure}
\usepackage{multirow}
\usepackage{ulem}
\usepackage{amsmath}
\usepackage{bm}
\usepackage{float}
\usepackage{booktabs}
\usepackage{threeparttable}
\usepackage{CJK}
\usepackage[whole]{bxcjkjatype} 
\usepackage{CJKutf8}
\usepackage[colorlinks,linkcolor=green, anchorcolor=green,citecolor=green]{hyperref}
\usepackage{ulem,soul}
\usepackage{xcolor}

\hypersetup{colorlinks=true,
	breaklinks=true,
	pdfstartview=Fit,
	linkcolor=blue,
	citecolor=blue,
	urlcolor=blue}

\begin{document}

\title[Type~II-P SN\,2018gj]{A binary merger product as the direct progenitor of a Type II-P supernova} 

\author{Zexi\,Niu$^{\dagger}$}
\affiliation{School of Astronomy and Space Science, University of Chinese Academy of Sciences, Beijing 100049, China}
\affiliation{National Astronomical Observatories, Chinese Academy of Sciences, Beijing 100101, China}

\author{Ning-Chen\,Sun$^{\dagger}$$^{*}$}
\affiliation{School of Astronomy and Space Science, University of Chinese Academy of Sciences, Beijing 100049, China}
\affiliation{National Astronomical Observatories, Chinese Academy of Sciences, Beijing 100101, China}
\affiliation{Institute for Frontiers in Astronomy and Astrophysics, Beijing Normal University, Beijing 102206, China}

\author{Emmanouil\,Zapartas$^{\dagger}$}
\affiliation{Institute of Astrophysics, Foundation for Research and Technology-Hellas, Heraklion GR-71110, Greece}
\affiliation{Physics Department, National and Kapodistrian University of Athens, Athens 15784, Greece}

\author{Dimitris\,Souropanis}
\affiliation{Institute of Astrophysics, Foundation for Research and Technology-Hellas, Heraklion GR-71110, Greece}

\author{Yingzhen\,Cui}
\affiliation{National Astronomical Observatories, Chinese Academy of Sciences, Beijing 100101, China}

\author{Justyn\,R.\,Maund}
\affiliation{Department of Physics, Royal Holloway, University of London, Egham TW20 0EX, United Kingdom}

\author{Jeff\,J.\,Andrews}
\affiliation{Department of Physics, University of Florida, Gainesville FL 32611, USA}
\affiliation{Institute for Fundamental Theory, Gainesville FL 32611, USA}

\author{Max\,M.\,Briel}
\affiliation{Département d’Astronomie, Université de Genève, Versoix CH-1290, Switzerland}
\affiliation{Gravitational Wave Science Center (GWSC), Université de Genève, Geneva CH1211, Switzerland}

\author{Morgan\,Fraser}
\affiliation{School of Physics, University College Dublin, Dublin 4 D04 P7W1, Ireland}

\author{Seth\,Gossage}
\affiliation{Center for Interdisciplinary Exploration and Research in Astrophysics (CIERA), Northwestern University, Evanston IL 60201, USA}
\affiliation{NSF-Simons AI Institute for the Sky (SkAI), Chicago IL 60611, USA}

\author{Matthias\,U.\,Kruckow}
\affiliation{Département d’Astronomie, Université de Genève, Versoix CH-1290, Switzerland}
\affiliation{Gravitational Wave Science Center (GWSC), Université de Genève, Geneva CH1211, Switzerland}

\author{Camille\,Liotine}
\affiliation{Center for Interdisciplinary Exploration and Research in Astrophysics (CIERA), Northwestern University, Evanston IL 60201, USA}
\affiliation{Department of Physics and Astronomy, Northwestern University, Evanston IL 60208, USA}

\author{Zhengwei\,Liu}
\affiliation{International Centre of Supernovae (ICESUN), Yunnan Key Laboratory of Supernova Research, Yunnan Observatories, Chinese Academy of Sciences (CAS), Kunming 650216, China}

\author{Philipp\,Podsiadlowski}
\affiliation{London Centre for Stellar Astrophysics, London, United Kingdom}
\affiliation{University of Oxford, St Edmund Hall, Oxford OX1 4AR, United Kingdom}

\author{Philipp\,M.\,Srivastava}
\affiliation{Electrical and Computer Engineering, Northwestern University, Evanston IL 60208, USA}
\affiliation{Center for Interdisciplinary Exploration and Research in Astrophysics (CIERA), Northwestern University, Evanston IL 60201, USA}
\affiliation{NSF-Simons AI Institute for the Sky (SkAI), Chicago IL 60611, USA}

\author{Elizabeth\,Teng}
\affiliation{Department of Physics and Astronomy, Northwestern University, Evanston IL 60208, USA}
\affiliation{Center for Interdisciplinary Exploration and Research in Astrophysics (CIERA), Northwestern University, Evanston IL 60201, USA}
\affiliation{NSF-Simons AI Institute for the Sky (SkAI), Chicago IL 60611, USA}

\author{Xiaofeng\,Wang}
\affiliation{Department of Physics, Tsinghua University, Beijing 100084, China}

\author{Yi\,Yang}
\affiliation{Department of Physics, Tsinghua University, Beijing 100084, China}

\author{Jifeng Liu$^{*}$}
\affiliation{National Astronomical Observatories, Chinese Academy of Sciences, Beijing 100101, China}
\affiliation{School of Astronomy and Space Science, University of Chinese Academy of Sciences, Beijing 100049, China}
\affiliation{Institute for Frontiers in Astronomy and Astrophysics, Beijing Normal University, Beijing 102206, China}
\affiliation{New Cornerstone Science Laboratory, National Astronomical Observatories, Chinese Academy of Sciences, Beijing 100012, China}

\begin{abstract}

Type~II-P supernovae (SNe\,II-P) are the most common class of core-collapse SNe in the local Universe and play critical roles in many aspects of astrophysics. Since decades ago theorists have predicted that SNe\,II-P may originate not only from single stars but also from interacting binaries. While $\sim$20 SN\,II-P progenitors have been directly detected on pre-explosion images, observational evidence still remains scarce for this speculated binary progenitor channel.
In this work, we report the discovery of a red supergiant progenitor for the Type~II-P SN\,2018gj. While the progenitor resembles those of other SNe\,II-P in terms of effective temperature and luminosity, it is located in a very old environment and SN\,2018gj has an abnormally short plateau in the light curve.
With state-of-the-art binary evolution simulations, we find these characteristics can only be explained if the progenitor of SN\,2018gj is the merger product of a close binary system, which developed a different interior structure and evolved over a longer timescale compared with single-star evolution.
This work provides the first compelling evidence for the long-sought binary progenitor channel toward SNe\,II-P, and our methodology serves as an innovative and pragmatic tool to motivate further investigations into this previously hidden population of SNe\,II-P from binaries.

\par{Keyword: Core-collapse supernovae, Type II supernovae, Massive stars evolution, Red supergiant stars}
\end{abstract}

\begin{center}
\begin{minipage}{0.9\textwidth}
$\dagger$ {These authors contributed equally to this work}\\
* Corresponding authors: sunnc@ucas.ac.cn, jfliu@nao.cas.cn
\end{minipage}
\end{center}

\maketitle

\section{Introduction}\label{intro.sec}

As the most common type ($\gtrsim$50\% \cite{Li2011}) of core-collapse supernovae (SNe) in the local Universe, Type~II-P refers to events whose spectra exhibit strong H features and whose light curves display a plateau phase for a few months powered by H recombination. They display a rich photometric and spectroscopic diversity, such as the plateau length, peak luminosity and expansion velocity \cite{Martinez2022a, Martinez2022b, Martinez2022c}. Understanding their origin is fundamental to unraveling the evolution of massive stars, the physics of their explosion, the lifecycle of the interstellar medium, the formation of compact objects, as well as the galactic chemical/mechanical feedback processes. The direct detection of $\sim$20 progenitors on pre-explosion images and analysis of their spectral energy distributions (SEDs) have confirmed that SNe\,II-P are exploding red supergiant (RSG) stars with massive H-rich envelopes \cite{Smartt2009, Rui2019, Niu2023, Xiang2024}.

It has now been well established that interacting binaries dominate the evolution of most massive stars \cite{Sana2012} and the origin of most H-poor SNe (i.e. Types IIb, Ib, Ic and Ibn; \cite{Podsiadlowski1992, 2004Maund, 2013Eldridge, 2017Yoon, Zapartas2017b, Sun2020, 2020Sun, Sun2022, Sun2023a, Niu2024, Zhao2025}). For the H-rich SNe\,II-P, theoretical works have also predicted for decades that they may originate not only from single-star evolution but also from interacting binaries \cite{Podsiadlowski1992}. The elaborate binary pathways towards SNe\,II-P may include (1) the coalescence of two stars (merger), (2) the partial stripping of the initially more massive star (mass donor), or (3) the accretion onto an initially less massive companion (mass gainer). Recent studies quantify that as many as 30\%--50\% of all SNe\,II-P may have binary progenitors \cite{Eldridge2018, Zapartas2019, 2024Dessart, Schneider2024,2025Ercolino}. With a very wide parameter space (initial masses and orbital separation) and often very complicated interaction between the two member stars, binaries also serve as a promising candidate to explain what drives the observational diversity of SNe II-P.

Compelling observational evidence, however, still remains elusive for this speculated binary progenitor channel toward SNe\,II-P. In both the single and binary channels, a SN\,II-P progenitor before core collapse will similarly appear as a RSG based on effective temperature and luminosity  \cite{Eldridge2018,Zapartas2019, Farrell2020, Schneider2024}. The SN spectra will neither exhibit any obvious signatures of binary interaction since most differences in the progenitor's chemical abundance profiles caused by material mixing during mass transfer will be erased in the subsequent stellar evolution. Survived binary companions are not expected to be found for the merger and mass gainer progenitors \cite{Zapartas2019,2025Zapartas_1} while the mass donor progenitors, which may possibly have survived binary companions, account for only $<$5\% of SNe\,II-P as they require fine-tuned orbital separations to retain sufficiently massive H-rich envelopes \cite{Eldridge2018}. Therefore, the binary progenitor channel for SNe\,II-P still remains an outstanding challenge in astrophysics.

A key prediction from theoretical population synthesis is that many merger and mass gainer progenitors have longer evolutionary timescales compared to single-star progenitors of equivalent RSG luminosity and He-core mass \cite{2016Schneider,Zapartas2017, Zapartas2021}. 
Thus, an innovative methodology to uncover the potential binary nature of SN\,II-P progenitors is to identify them in older stellar environments than those consistent with single-star evolution \cite{Zapartas2021,Bostroem2023}. 
In addition, binary progenitors, depending on the mass-transfer history, may have very different core-to-envelope mass ratios compared to single-star progenitors \cite{Schneider2024,2020Schneider}, and this can lead to unusual plateau lengths in the resultant SN light curves compared with those from single-star progenitors \cite{Eldridge2018}.

In this work, we report the identification of a RSG progenitor for the Type\,II-P SN\,2018gj. We provide compelling observational evidence for its origin from an interacting binary system and reveal its pre-SN evolution with state-of-the-art binary evolution simulations. This study confirms that interacting binaries are indeed a viable progenitor channel toward SNe\,II-P and establishes a successful methodology that can motivate future investigations to uncover a previously hidden population of SNe\,II-P from binaries.

\section{Progenitor Detection}\label{detection.sec}

\nocite{Teja2023}

\begin{figure*}[htbp]
    \centering
    \includegraphics[width=0.85\textwidth]{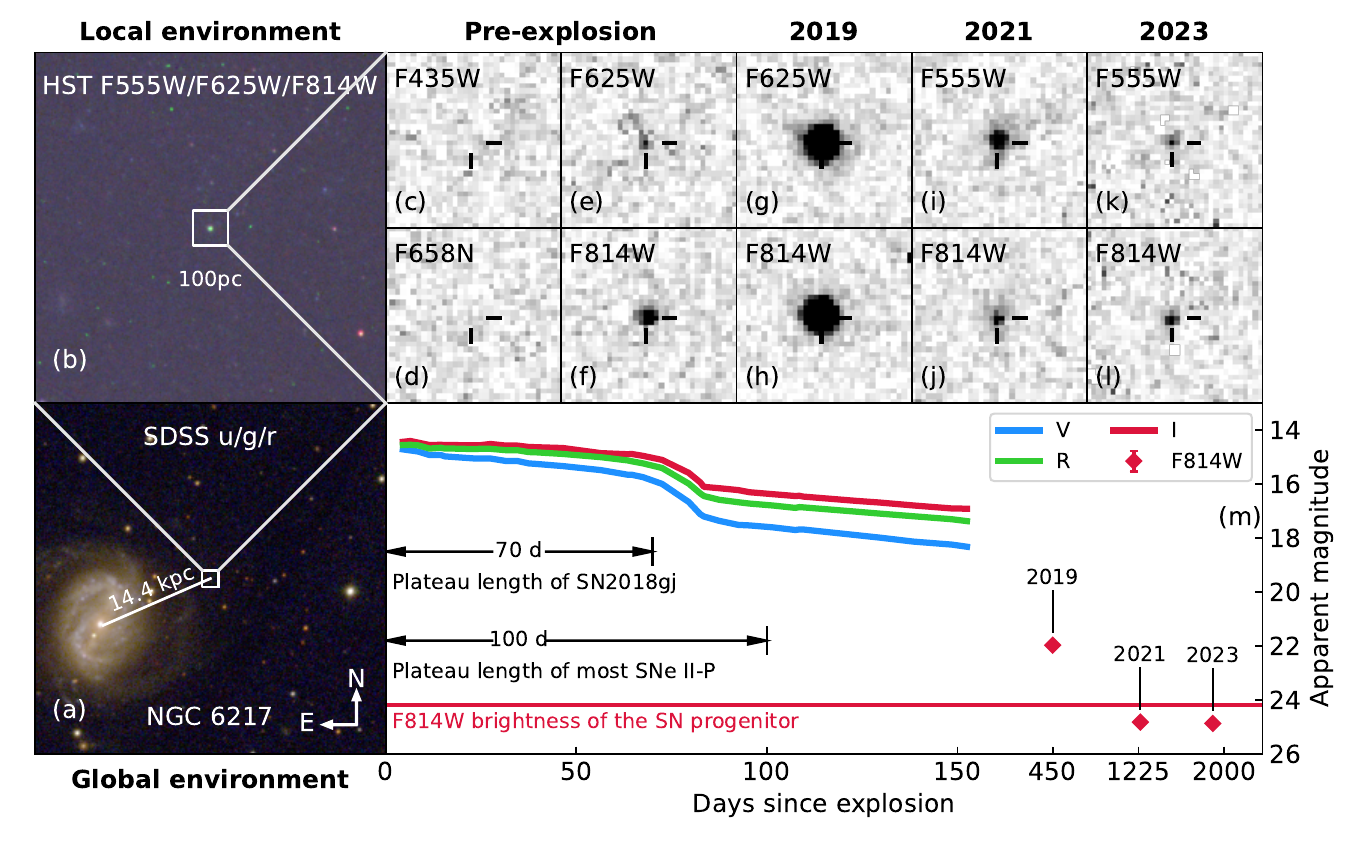}
    \caption{Direct pre-explosion detection of SN\,2018gj's progenitor and the photometric evolution of SN\,2018gj. (a) SDSS \textit{u/g/r} composite image of NGC\,6217, where SN\,2018gj is located in the far outskirt. (b) HST F555W/F625W/F814W composite image of the local SN environment, which is a very sparse region without any prominent features of recent star formation. (c--l) HST images taken before and after the explosion of SN\,2018gj. The crosshairs indicate the SN position. (m) Light curves of SN\,2018gj from the literature \cite{Teja2023} and late-time photometry from the HST observations. SN\,2018gj has a very short plateau length of 70\,days in contrast to 100\,days for most SNe\,II-P. In 2021 and 2023, the F814W brightness becomes much fainter than the pre-explosion level, suggesting that the SN progenitor has disappeared after explosion.}
    \label{fig:fig1}
\end{figure*}

SN\,2018gj is a Type\,II-P event located in the outskirt of the nearby galaxy NGC\,6217 (Fig.\,\ref{fig:fig1}a) with a heliocentric redshift of $z$ = 0.004540. 
The site of SN\,2018gj was observed by the HST several times before and after its explosion with the Wide Field Channel (WFC) of the Advanced Camera for Surveys (ACS) and the Ultraviolet-Visible channel (UVIS) of the Wide Field Camera 3 (WFC3). A full list of these observations is provided in Table\,\ref{tab:18gj_HST}. We retrieved the images that have been flat-fielded and corrected for charge transfer efficiency (i.e. \texttt{*\_flc.fits}) from the Barbara A. Mikulski Archive for Space Telescopes (MAST; \url{https://mast.stsci.edu}). For the post-explosion images, we manually combined the exposures with the \textsc{astrodrizzle} package for better cosmic-ray removal by setting \texttt{driz$\_$cr$\_$grow = 3}. The pre-explosion observations adopted an unusually large dithering distance of 2048 pixels and an exposure at only one of the dithering positions covered SN\,2018gj; for these images, we removed the cosmic rays with Laplacian edge detection \cite{LACosmic.ref,curtis_mccully_2018_1482019}.

We used the \textsc{dolphot} package \cite{dolphot.ref} for point-spread-function (PSF) photometry with parameters recommended in the user manual. On the post-explosion images taken in 2019, 2021 and 2023 (Fig.\,\ref{fig:fig1}g--l), a continuously fading source is detected near the reported coordinates of SN\,2018gj, ($\alpha$, $\delta$)$_{\rm J2000}$ = (16:32:02.300, +78:12:40.9). We identified this fading source as the late-time radiation of SN\,2018gj. We then performed differential astrometry between the pre- and post-explosion images taken in 2019 with 10 common stars, which allowed us to determine the SN position on the pre-explosion images with an accuracy of 0.83 pixels. A point source was significantly detected on the pre-explosion F625W and F814W images (Fig.\,\ref{fig:fig1}e and f) at the derived SN position with an offset of only 0.16 pixels, much smaller than the astrometric uncertainty. 
On the pre-explosion F435W and F658N images (Fig.\,\ref{fig:fig1}c and d), no source is detected at the SN position at a $3\sigma$ level. We determined the detection limit by inserting artificial stars with different magnitudes at the SN position; the detection limit corresponds to the magnitude where the $3\sigma$ source detection probability falls to 50\%.
The last column of Table\,\ref{tab:18gj_HST} lists the photometry of this source and SN\,2018gj at late times.

We suggest that this pre-explosion point source corresponds to the progenitor of SN\,2018gj. The SN is located in a very sparse area, so chance alignment with an unrelated celestial object is very unlikely, and we estimated a chance-alignment probability of only 0.004\% based on the surface density of sources in this area. The detected progenitor is neither due to an unrejected cosmic ray, since a randomly positioned cosmic ray will coincide with SN\,2018gj within the positional error circle with a very low probability of 0.012\%. Furthermore, the \texttt{SHARPNESS} parameter returned by \textsc{dolphot} is $-$0.007, very close to zero; this suggests that the morphology of the detected progenitor matches the PSF very well, while cosmic rays are usually much sharper than the PSF. The F814W brightness of SN\,2018gj in 2021 and 2023 becomes much fainter than the progenitor, suggesting that it has disappeared after the SN explosion. We notice that although the F814W brightness in 2021 and 2023 is higher than a simple extrapolation of the light curve tail, this could be attributed to a light echo or CSM interaction, which are commonly observed in many SNe (e.g., Refs. \cite{2015VanDyk,2020Sun}).

SN~2018gj was also observed with the Spitzer/IRAC at 3.6 and 4.5 $\mu$m in 2009 and 2019. We retrieved the post–basic calibrated data (PBCD) images from the Spitzer Heritage Archive\footnote{\url{https://sha.ipac.caltech.edu}}. The SN is clearly visible in both bands in 2019, but no pre-explosion counterpart can be detected at the SN position in the 2009 images. Differential astrometry with 16 common stars is applied, achieving a precision of 0.06/0.18 pixels for 3.6/4.5-$\mu$m band. Based on the local background flux densities around the SN position, we estimated 3-$\sigma$ detection limits for the progenitor of [3.6] $>$ 16.5~mag and [4.5] $>$ 15.9~mag. Zeropoints and conversion factors between flux density and Vega magnitudes were adopted from the IRAC Instrument Handbook\footnote{\url{https://irsa.ipac.caltech.edu/data/SPITZER/docs/irac/iracinstrumenthandbook}}. The model SEDs adopted in our analysis (see below) predict pre-explosion magnitudes of about 21~mag at [3.6]/[4.5] for the progenitor system, much fainter than the detection limits.
While these limits are not sufficiently deep to provide meaningful constraints on progenitor SED models, we include them here for completeness.

\begin{table*}
\small
\centering
\caption{HST observations and photometry of SN\,2018gj. \label{tab:18gj_HST}}
\begin{tabular}{ccccccc}
\hline 
Date & Epoch$^{a}$ & Instrument & Filter & Exposure & Program & Magnitude$^{c}$ \\
 & (day) & & & Time (s) & ID & (mag) \\
\hline
2009-06-13 & $-$3133 & ACS/WFC & F435W & 500$^b$ & 11371 & $>$28.5 \\
2009-06-13 & $-$3133 & ACS/WFC & F625W & 400$^b$ & 11371 & 26.10 (0.23) \\
2009-07-08 & $-$3108 & ACS/WFC & F658N & 500$^b$ & 11371 & $>$24.9 \\
2009-07-08 & $-$3108 & ACS/WFC & F814W & 400$^b$ & 11371 & 24.19 (0.05) \\
2019-05-16 & 491 & WFC3/UVIS & F625W  & 1305 & 15151 & 22.51 (0.01) \\
2019-05-16 & 491 & WFC3/UVIS & F814W  & 1305 & 15151 & 21.97 (0.02) \\  
2021-06-27 & 1264 & WFC3/UVIS & F555W  & 710 & 16179 & 25.07 (0.06) \\  
2021-06-27 & 1264 & WFC3/UVIS & F814W  & 780 & 16179 & 24.82 (0.09) \\  
2023-02-27 & 1874 & ACS/WFC & F555W  & 760 & 17070 & 25.91 (0.18) \\  
2023-02-27 & 1874 & ACS/WFC & F814W  & 780 & 17070 & 24.87 (0.07) \\ 
\hline
\end{tabular}
\begin{flushleft}
$^a$ Relative to an estimated explosion date of 2018-04-10 \cite{Teja2023}. \\
$^b$ The effective exposure time of these observations, as listed here, is half of the total exposure time since the SN position is covered by only one of the two dithered exposures. \\
$^c$ All magnitudes reported in this paper are in the Vega magnitude system. \\
\end{flushleft}
\end{table*}

\nocite{marcs.ref}
\nocite{bpass.ref}

\begin{figure*}[htbp]
    \centering
    \includegraphics[width=0.8\textwidth]{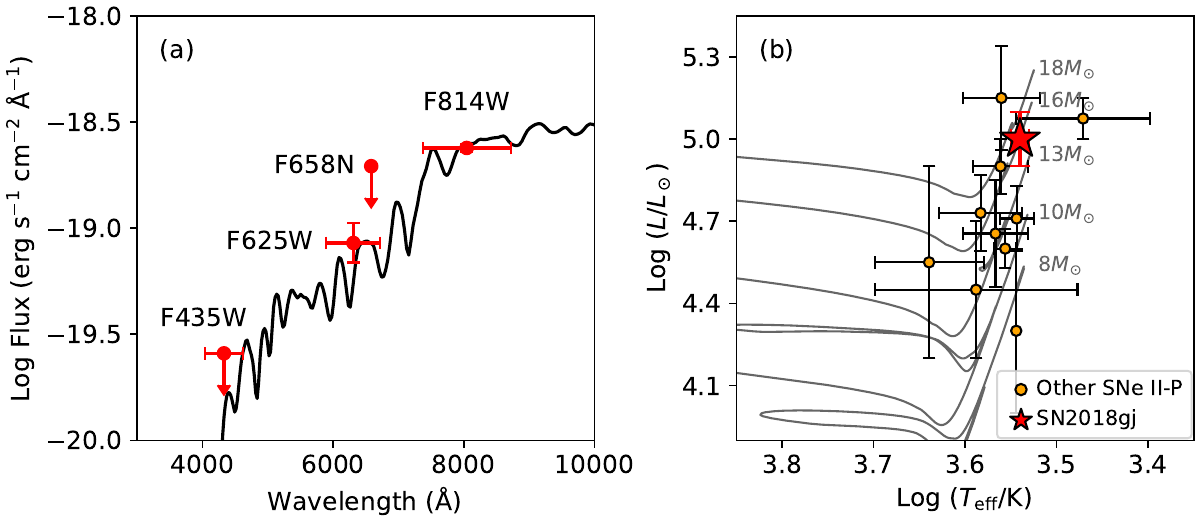}
    \caption{Progenitor properties derived from the direct detections.(a) SED of the SN progenitor and the best-fitting \textsc{marcs} \cite{marcs.ref} theoretical spectrum, consistent with being a luminous RSG. (b) Position of SN\,2018gj's progenitor on the HRD in comparison with other confirmed SNe\,II-P progenitors and the \textsc{bpass} \cite{bpass.ref} single-star evolutionary tracks for solar metallicity.}
    \label{fig:fig2}
\end{figure*}

Fig.\,\ref{fig:fig2}a displays the SED of the SN progenitor. We fit the observed SED of SN\,2018gj's progenitor with the \textsc{marcs} model atmospheres with a solar metallicity, assuming a microturbulent velocity of 5\,km\,s$^{-1}$ and surface gravity log($g$) = 0\,dex (broad-band photometry is not sensitive to the values of these two parameters).  Values of distance, metallicity and line-of-sight reddening are discussed in the Supplementary Materials. The effective temperature and bolometric luminosity were left as free parameters to be fitted. We calculated synthetic magnitudes by convolving the model atmospheres with the HST filters' response curves, which were then compared with the observed magnitudes or detection limits. We derived the posterior probability distributions with a Markov-Chain Monte-Carlo method and found an effective temperature of log($T_{\rm eff}$/K) = 3.54 $\pm$ 0.01 and a luminosity of log($L/L_\odot$) = 5.0 $\pm$ 0.1, corresponding to a radius of 875\,$\pm\,140$~$R_\odot$. The uncertainties were propagated from those in the photometry, distance and reddening. Fig.\,\ref{fig:fig2}b plots the position of the progenitor on the Hertzsprung-Russell diagram (HRD); other confirmed SN\,II-P progenitors are overlaid for comparison. 

For RSGs, the final He-core mass and final luminosity follow a tight relation \cite{Farrell2020}: 
\begin{equation}
{\rm log}(M_{\rm He}/M_\odot) = 0.659 \times {\rm log}(L/L_\odot) -2.630;
\label{eq.Hecore}
\end{equation}
therefore, the progenitor of SN\,2018gj had a final He-core mass of $M_{\rm He}$ = 4.6\,$\pm$\,0.7\,$M_\odot$. 
Assuming single-star evolution, this corresponds to an initial mass of $M_{\rm ini}\,=\,13$--$16\,M_{\odot}$ and an age of log($t$/yr)\,=\,7.1--7.3 based on the \textsc{bpass} \cite{bpass.ref} and \textsc{binary\_c} \cite{Izzard+2004} single-star evolutionary tracks. The results from \textsc{posydon} single star models  \cite{posydon.ref,posydonv2.ref} yield $M_{\rm ini}\,=\,$12--15$\,M_{\odot}$ and log($t$/yr)$\,=\,$7.2--7.4. These three models are mainly used in the following analysis.
These values are also consistent with results based on other stellar evolutionary models (i.e. \textsc{kepler}, \textsc{hoshi} \cite{2025Fang}). 

Another probe of the progenitor state, especially its core mass, comes from the [O\textsc{i}] $\lambda\lambda$6300, 6363 doublet flux at nebular phase, which for SN\,2018gj  is in good agreement with the model spectrum of an initial progenitor mass of 15~$M_{\odot}$ \cite{Teja2023}. This agreement has also been verified through a separate analysis (\cite{2025Fang}, private communication). 
This independent measurement is consistent with the single-star equivalent initial mass from our progenitor detection. 

Some SNe\,II-P progenitors are subject to circumstellar extinction and brightness variability due to pulsations, which may influence the luminosity/mass estimate based on direct progenitor detections.  The color of SN\,2018gj progenitor, when corrected for (Galactic and host) interstellar extinction, agrees well with that of a RSG and does not show any obvious color excess arising from circumstellar extinction. Besides, no significant dust emission can be detected at the SN position on the 3.6- and 4.5-$\mu m$ pre-explosion Spitzer/IRAC images.  Therefore, it seems unlikely that the derived progenitor 
Pulsational variability of the progenitor's luminosity \cite{2018Morozova, Goldberg+2020, 2025Bronner} has been inferred in various event, such as SN\,2017eaw \cite{2019VanDyk}, SN\,2023ixf \cite{2024Zimmerman,Niu2023,2025Laplace}, SN\,2024ggi \cite{Xiang2024}, and SN\,2025pht \cite{2025Kilpatrick}, which may influence the inferred pre-SN core mass estimate. 
In an extreme case that SN~2018gj is analogous to the case of SN~2023ixf with which it has similar progenitor luminosity, its value may vary by up to 0.25~dex \cite{2025Laplace}, with pulsations being the dominant source of uncertainty. This would imply a lower limit of the final He-core mass of $M_{\rm He}\approx3.2~M_\odot$, which we further discuss in the context of our results in Section~\ref{evidence_for_binary.sec}.

\section{Single or binary evolution?}\label{singe_or_binary.sec}

While the directly detected progenitor of SN\,2018gj resembles those of other SNe\,II-P in terms of effective temperature and luminosity, SN\,2018gj exhibits two distinctive observational characteristics.
(1) In contrast to most core-collapse SNe that are associated with active star-forming regions, SN\,2018gj is located in a very sparse area as far as 14.4~kpc from the host-galaxy center and without any obvious signs of recent star formation.
(2) While the peak luminosity, color, and spectral evolution are all typical for SNe\,II-P, SN\,2018gj displays a remarkably short plateau of only 70\,$\pm$\,2\,days \cite{Teja2023} compared with an average plateau length of $\sim$100\,$\pm$\,20\,days for most SNe\,II-P \cite{Valenti2016, Hicken2017} (Fig.\,\ref{fig:fig1}m). In this section, we shall investigate the origin of SN\,2018gj based on the direct progenitor detection combined with these two characteristics.

\subsection{Environment and age}\label{age.sec}

We analyzed the environment of SN\,2018gj with the surrounding stellar populations detected on the post-explosion HST/WFC3 images, which have a better spatial resolution and deeper detection limits than those acquired by ACS. In each band (F555W, F625W and F814W), we considered a star to be significantly detected if the parameters reported by \textsc{dolphot} met the following criteria:
\begin{itemize}
    \item type of source, \texttt{TYPE} = 1;
    \item signal-to-noise ratio, \texttt{SNR} $\geq$ 3;
    \item photometry quality flag, \texttt{FLAG} $\leq$ 2;
    \item source sharpness, $-$0.5 $<$ \texttt{SHARPNESS} $<$ 0.5;
    \item source crowding, \texttt{CROWDING} $<$ 1.
\end{itemize}
These criteria excluded bad detections, extended sources and cosmic-ray residuals, leaving only well-measured point sources. We also used artificial star tests to estimate the detection limit in each band; an artificial star was considered to be successfully recovered if it was found within 1 pixel of the inserted position and its \textsc{dolphot} parameters met all the above criteria. The detection limit in a filter was characterized by $m^{\rm lim} \pm \sigma^{\rm lim}_m$, where $m^{\rm lim}$ is the magnitude where the detection probability falls to 50\% and $\sigma^{\rm lim}_m$ is its ``uncertainty", i.e. the magnitude range where the recovery rate declines from 68\% to 32\%.

\begin{figure*}[htbp]
    \includegraphics[width=0.8\textwidth]{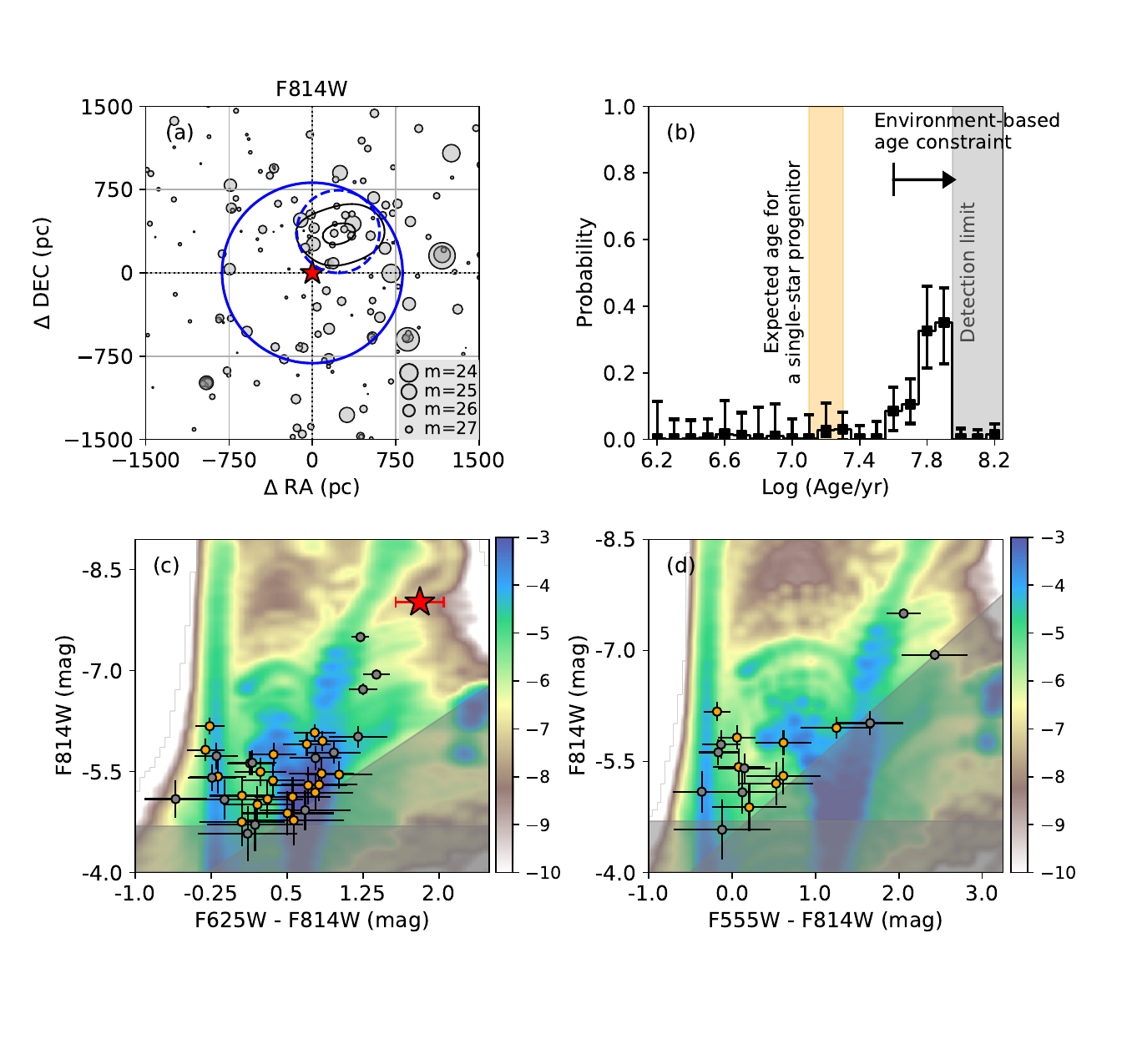}
    \caption{Estimation of the stellar age for SN\,2018gj's progenitor.
    (a) Map of stars detected on the HST/WFC3/F814W images in the environment of SN\,2018gj with the symbol size reflecting their magnitudes.
   (b) The stellar age distribution in the SN local environment derived from the CMD fitting. (c,d) CMD of the resolved stellar populations inside the $R_{\rm od}$ region (gray) and inside the $R_{\rm env}$ region but outside the $R_{\rm od}$ region (orange) in comparison with the SN progenitor (red point). 
   The error bars are the photometric uncertainties and the grey-shaded area corresponds to the detection limits. The background color scale shows the normalized probability density distribution for the environmental sources simulated with the \textsc{bpass} binary population models.
    }
    \label{fig:fig3}
\end{figure*}

Fig.\,\ref{fig:fig3}a shows the map of stars detected in the SN vicinity. It is obvious that SN\,2018gj is located in a very sparse region with a very uniform stellar distribution, except for a stellar surface over-density to its northwest. This over-density may arise from the random Poisson noise of a uniform distribution; it is also possible that the over-density is the relic of an old star-forming complex that has almost dispersed. 
We define a small circular region with a radius of 380\,pc covering the over-density (dashed circle, hereafter $R_{\rm od}$); we also define a large circular region with a radius of 850\,pc centered on the SN position and encompassing the over-density (solid circle, hereafter $R_{\rm env}$). There are 131 sources within the $R_{\rm env}$ region and 42 sources inside the $R_{\rm od}$ region. Their Color-magnitude diagrams (CMDs) are shown in Fig.\,\ref{fig:fig3}c and d. There is no significant difference between those inside and outside the overdensity. This suggests that the over-density stars should have very similar ages to the other stars in the SN vicinity.

We use stars within the above-defined $R_{\rm env}$ region to derive the stellar age distribution in the SN environment. This is performed by fitting their CMDs with a hierarchical Bayesian approach \cite{Maund2016, Sun2021} as detailed in the Supplementary Materials. The derived stellar age distribution is shown in Fig.\,\ref{fig:fig3}b and the best-fitting model CMDs are displayed in Fig.\,\ref{fig:fig3}c and d. Most stars were formed at log($t$/yr)~$\geq$~7.60~$\pm$~0.05 and very few, if any, were formed at log($t$/yr)~=~7.1--7.3, the progenitor age of SN\,2018gj expected from single-star evolution.

In previous studies, much smaller regions (i.e. 100--200~pc, \cite{Maund2016,Sun2023a}) were usually used to study SN environments such that stars within the regions are coeval with the SN progenitor and an accurate age can be derived for the progenitor.
For SN\,2018gj, however, using a large region should not be a problem since we only found a lower age limit for the SN progenitor -- if there are no young stars within 850\,pc, there are no young stars within 100--200~pc. Besides, as discussed above, this region is very sparse and uniform, without obvious age variation between different positions. This is not like the young star-forming regions with high spatial variations. Therefore, the defined $R_{\rm env}$ region is representative for the environment of SN\,2018gj. We have also conducted tests using 28 sources within 400\,pc, and the fitted stellar age distribution does not change significantly except for larger uncertainties. Using a large region can include a larger number of stars and reduce the random errors in the derived stellar age distribution. 

It is also worth mentioning that the defined region is large enough to include the birthplace of the progenitor even if it is a walkaway or runaway star ejected from a prior binary system, which can travel an average distance of $\sim$100~pc during its short lifetimes \cite{2019Renzo,2025Wagg}. 
The birthplace may lie outside the defined region if the progenitor was a hypervelocity star ejected from the host galaxy's center. However, we deem this scenario highly unlikely due to the low ejection rate \cite{2018Boubert,2022Marchetti}. Moreover, given the distance to the galactic center and the lifetime of a single star with $M_{\rm He} = 4.6 \pm 0.7\,M_\odot$, even under an extreme ejected velocity of 1000~km s$^{-1}$, the progenitor would need to have been ejected very early during its main sequence (MS), when the progenitor was compact and had low mass-loss rate. This is in contradiction to the low-mass H-rich envelope for SN\,2018gj (see below).

The above analysis is based on the stellar population around the SN. 
The possibility of an underlying stellar cluster at the SN position can be ruled out. Given the deep detection limit in the F435W band on the pre-explosion image, the initial mass of any potential stellar cluster younger than 50~Myr (the lifetime of a 8~$M_{\odot}$ star) would be less than 50~$M_{\odot}$. However, such a very low‑mass cluster is expected to host only 0.04 massive stars. Therefore, it is very unlikely that the progenitor of SN\,2018gj is associated with a star cluster at the SN position. 

~\\
\subsection{Light curve and final H-envelope mass}\label{H_env.sec}

The short plateau length in the light curve points to an abnormally low-mass H-rich envelope for the progenitor of SN\,2018gj. Ref.\,\cite{Teja2023} estimated the final H-envelope mass by carrying out hydrodynamical modelling with \textsc{mesa} \cite{mesa.ref1} and \textsc{stella} \cite{stella.ref1} (their Section\,5). They tried two sets of models with progenitor initial masses of $M_{\rm ini}$ = 13\,$M_\odot$ and 19\,$M_\odot$, respectively, and changed the final H-envelope mass by arbitrarily tuning the wind mass-loss rate. They found that the pseudo-bolometric light curve and the Fe\,\textsc{ii} $\lambda$\,5169\,\AA\ velocity evolution can be fitted with an explosion energy of 4~$\times$~10$^{50}$~erg, an initial progenitor mass of $M_{\rm ini} = 13\,M_\odot$ and a final H-envelope mass of $M_{\rm H}$ = 3.26\,$M_\odot$ which was achieved by increasing the default wind mass-loss rate by a factor of 5.

We also carried out a similar analysis by employing MESA code (version 15140) for an initial progenitor mass of $M_{\rm ini} = 15~M_\odot$ (based on the MESA test suite \texttt{15M$\_$dynamo}), which is more consistent with our progenitor detection. We show models with different final H-envelope masses in the Supplementary materials. 
The best-fitting model has a final H-envelope mass of $M_{\rm H}$ = 3.5~$\pm$~0.5~$M_\odot$, achieved by scaling the default wind mass-loss rate by a factor of 3.0--3.5. 
This final H-envelope mass is adopted for further analysis in this work. It is very similar to that of \cite{Teja2023}, which is reasonable since the SN light curve shape is very sensitive to H-rich envelope but depends very little on the He-core mass \cite{2019Dessart}.

Interaction between the SN ejecta and circumstellar material (CSM) may affect the light curve and cause uncertainties in measuring the final H-envelope mass \citep{Davies+2022,Morozova+2018}. For SN\,2018gj, no narrow emission lines from ejecta-CSM interaction can be seen in the spectra \cite{Teja2023}. In Ref.\,\cite{Teja2023}, a CSM of $<$0.15\,$M_{\odot}$ was incorporated to fit the early pre-peak light curve. However, this CSM mass is very small and has very little effect on the plateau after peak. Therefore, the possible effect of CSM on measuring the final H-envelope mass can be neglected.

In addition, a recent study by Ref.\,\cite{2025Fangql} presents scaling relations between SN observables and progenitor properties, considering the partial stripping of the H-rich envelope. Degeneracies in the scaling relations can be further broken \cite{2020Goldberg} given the progenitor radius 
(875\,$\pm\,140$~$R_\odot$, inferred in this work), $^{56}$Ni mass from the light-curve tail (0.025--0.03~$M_\odot$, \cite{Teja2023}), and explosion energy from the bolometric light curve and Fe\,\textsc{ii} $\lambda$\,5169\,\AA\ velocity evolution (4~$\times$~10$^{50}$~erg, \cite{Teja2023}). With these parameters, the plateau length corresponds to a final H-envelope mass of $M_{\rm H}$ = 3.00--$3.68~M_\odot$ based on the scaling relations. This is also consistent with our derived value.

We notice that Ref.\,\cite{2024Utrobin} presented a different hydrodynamical modelling and derived a significantly larger SN ejecta mass of 23~$M_{\odot}$, corresponding to an initial progenitor mass of 29~$M_{\odot}$. 
Their results are significantly larger than that inferred from SN nebular spectroscopy and direct detection of the progenitor. Furthermore, the explodability of such a massive progenitor also warrants consideration \cite{2014Sukhbold}; even if such a star were to explode, it would manifest itself as a H-poor SN rather than a H-rich SN due to its powerful stellar winds (e.g., Refs. \cite{2003Heger,2007Crowther}).

~\\
\subsection{Evidence for binary origin}\label{evidence_for_binary.sec}

\begin{figure*}[htbp]
    \centering
    \includegraphics[width=0.8\textwidth]{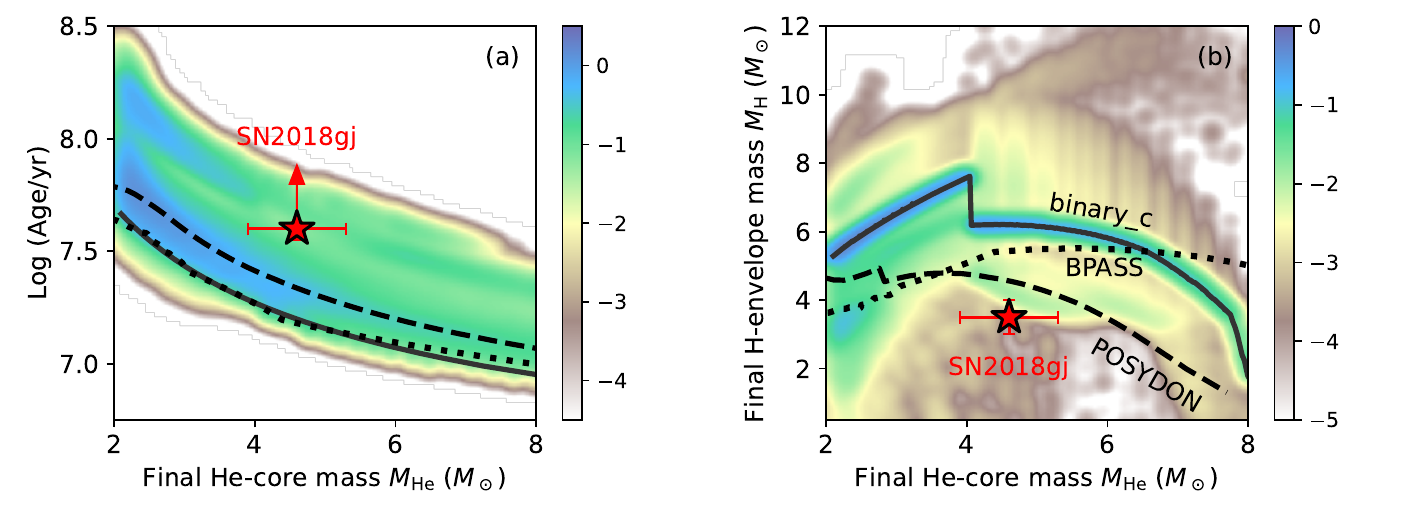}
    \caption{Final He-core mass ($M_{\rm He}$), final H-envelope mass ($M_{\rm H}$) and age of SN\,2018gj's progenitor. For comparison, the solid/dotted/dashed black lines show the relations between the parameters for single-star progenitors from \textsc{binary\_c}/\textsc{bpass}/\textsc{posydon} and the color scales show the probability density distributions for binary progenitors as predicted by \textsc{binary\_c} \cite{Zapartas2017}. The probability densities are on logarithmic scales and the color bars are in units of $M_\odot^{-2}$ (panel a) or $M_\odot^{-1}$\,dex$^{-1}$ (panel b).}
    \label{fig:fig4}
\end{figure*}

With the final He-core mass, final H-envelope mass and environment-based age constraint, we were able to determine whether a binary progenitor is more favored than a single-star progenitor for SN\,2018gj.
This is achieved by comparing the observables with theoretically predicted values for single and binary progenitors (Fig.\,\ref{fig:fig4}). For the single progenitors, we show the models from \textsc{binary\_c}, \textsc{posyndon} or \textsc{bpass} while for the binary progenitors we show the results from the \textsc{binary\_c} population synthesis \cite{Izzard+2004,Izzard2004,Izzard+2006,Izzard+2009,Zapartas2017}. While \textsc{binary\_c} has a limited accuracy based on analytical formulae of pre-computed single stellar models, it has the advantage of being able to rapidly calculate all possible binary scenarios and their contribution in a homogeneous way, implementing a large number of binary models on a dense grid of parameters that have already been tested in previous dedicated studies on the age and pre-SN properties of binary SN progenitors \cite{Zapartas2017,Zapartas2017b,Zapartas2019,Zapartas2021}.

As shown in Fig.\,\ref{fig:fig4}, SN\,2018gj's progenitor is a clear outlier from single-star models (be it \textsc{binary\_c}, \textsc{posyndon} or \textsc{bpass}). Assuming a single star, the progenitor should have an initial mass of $M_{\rm ini}$~=~13--16~$M_\odot$ and an age of log($t$/yr)~=~7.1--7.3. The environment of SN\,2018gj, however, is old and sparse, where most stars were formed at log($t$/yr)~$\geq$~7.60~$\pm$~0.05. Moreover, the progenitor has an abnormally low final H-envelope mass of only $M_{\rm H}$~=~3--4~$M_\odot$ while those for most SNe\,II-P lie in the range of $M_{\rm H}$~$\sim$~6--8~$M_\odot$ \cite{2025Fangql}. Given that the (single-star equivalent) progenitor mass and the roughly solar metallicity of SN\,2018gj are quite normal for SNe\,II-P \cite{Smartt2009}, the stripped envelope is very unlikely to be a consequence of a mass- and metallicity-dependent wind in single-star evolution. 

Quantitatively, we use a statistical Bayesian approach to distinguish single-star and interacting-binary progenitor channels. This can be regarded as a model-selection problem, in which
\begin{equation}
\dfrac{P({\rm BIN} | D)}{P({\rm SIN} | D)} = \dfrac{P(D | {\rm BIN})}{P(D | {\rm SIN})} \dfrac{P({\rm BIN})}{P({\rm SIN})},
\end{equation}
where $D=(M_{\rm H}, M_{\rm He}, {\rm log}(t))$ is the observable, $P({\rm BIN})$ and $P({\rm SIN})$ are the prior probabilities for the single and binary progenitor models, $P({\rm BIN} | D)$ and $P({\rm SIN} | D)$ are the posterior probabilities for the models given the observational data. 
The ratio of the posterior is the prior odds multiplied by the ratio of the evidence; the latter quantity is also regarded as the Bayes factor (BF) in favor of binary-progenitor model. 
Population synthesis has predicted that roughly 30--50$\%$ of Type~II SNe progenitors have a history of binary interaction (i.e. \cite{Podsiadlowski1992,Zapartas2019,2025Ercolino}), and we conservatively apply theoretical priors of $P({\rm BIN})=0.4$ and $P({\rm SIN})=0.6$.
In this case,
\begin{equation}
P_{\rm BIN}=\frac{2\times BF}{3+2\times BF}
\end{equation}
describes the relative probability of binary-progenitor model.

The Bayesian model evidences, $P(D | {\rm mod})$ with mod = SIN or BIN, can be calculated by marginalizing the model parameters:
\begin{equation}
\label{eq:evidence}
P(D | {\rm mod}) = \int{p(D | \theta, {\rm mod}) p(\theta | {\rm mod})d\theta},
\end{equation}
where $\theta$ = $M_1$ (i.e. stellar initial mass) or ($M_1$, $M_2$, $P$) (i.e. the initial primary mass, secondary mass and binary orbital period) in the single and binary progenitor models, respectively. 
$p(\theta | {\rm mod})$ is the prior probability distribution for the model parameters and $p(D | \theta, {\rm mod})$ is the likelihood. Given $\theta$, \textsc{binary\_c} has already synthesized a large population of single and binary SN progenitors with predicted properties of $\mu$ = ($M_{\rm H}^{\rm mod}, M_{\rm He}^{\rm mod}, {\rm log}(t)^{\rm mod}$). Therefore, Eq.\,\eqref{eq:evidence} is equivalent to:
\begin{equation}
\label{eq_model_i}
P(D | {\rm mod}) = \sum_{i}^{N_{{\rm mod}}} p_i \times
P(D | {\rm mod}_i),
\end{equation}
where mod$_i$ is the $i$-th model and $p_i$ is its weight in the synthesized population. The $i$-th model has predicted values of $\mu_i$ = ($M_{{\rm H}, i}^{\rm mod}, M_{{\rm He}, i}^{\rm mod}, {\rm log}(t)_i^{\rm mod}$).

With the derived values of $M_{\rm H}$ and $M_{\rm He}$ and the environment-based age constraint for SN\,2018gj, Eq.\,\eqref{eq_model_i} can be further expressed as:
\begin{equation}
\label{eq_pop_k}
P(D | {\rm mod}) = \sum_{i}^{N_{{\rm mod}}} p_i \times
\left[ \sum_{k}^{N_{{\rm pop}}} w_k \times P(D_k | {\rm mod}_i)\right],
\end{equation}
where $w_k$ is the weight of the $k$-th model population and $D_k$ = ($M_{\rm H}$, $M_{\rm He}$, log($t_k$)) with log($t_k$) being the age of this population (see Section~\ref{age.sec}). We use Gaussian likelihood for $P(D_k | {\rm mod}_i)$:
\begin{multline}
p(D_k | {\rm mod}_i) = (2\pi)^{-3/2}|\Sigma|^{-1/2} \\
\times \exp \left[ -\frac{1}{2} (D_k - \mu_i)^T \Sigma^{-1} (D_k - \mu_i) \right].
\end{multline}
We consider both the measurement and model uncertainties of the three observables and the error covariance matrix takes the form
\begin{multline}
\label{eq:Cov}
\Sigma =
\begin{bmatrix}
\sigma_{M_{\rm H, obs}}^2 & 0 & 0 \\
0 & \sigma_{M_{\rm He, obs}}^2 & 0 \\
0 & 0& \sigma_{\log(t), \rm obs}^2
\end{bmatrix}
+ \\
\begin{bmatrix}
\sigma_{M_{\rm H, mod}}^2 & 0 & 0 \\
0 & \sigma_{M_{\rm He, mod}}^2 & \rho \sigma_{M_{\rm He, mod}}\sigma_{\log(t), \rm mod} \\
0 & \rho \sigma_{M_{\rm He, mod}}\sigma_{\log(t), \rm mod} & \sigma_{\log(t), \rm mod}^2
\end{bmatrix}.
\end{multline}

The former term in Eq.\,\eqref{eq:Cov} corresponds to the measurement uncertainties, in which we use half of the age bin size (i.e. 0.05~dex) as the measurement uncertainty for $\sigma_{\log(t), \rm obs}$. 
The latter term corresponds to the model uncertainties. 
The uncertainty of the final H-envelope mass was  adopted as $\sigma_{M_{\rm H, mod}}$ = 20\% $M_{\rm H, mod}$, primarily due to the highly uncertain mass-loss prescriptions especially during the RSG phase \cite{2020Beasor,2025Zapartas}. This can be also observed from the discontinuity of $M_{\rm H}$ for the single-star track at $M_{\rm He}$ = 4~$M_\odot$ (Fig.\,\ref{fig:fig4}b), which is due to different mass-loss prescriptions used for different progenitor mass regimes.
Material mixing induced by stellar rotation and convective overshooting can transport the H-rich material from the envelope to the core; this process adds new fuel for nuclear burning and results in larger final He-core mass and longer stellar lifetime \cite{2012Ekstrom,2014Sukhbold,2020Kaiser}. 
For the uncertainty in the final He-core mass, we used the difference between typical assumptions in convective core overshooting such as the assumption in Ref.\,\cite{2011Brott} (implemented in \textsc{posydon} and calibrated for massive stars) and of lower values implemented in the models of  \textsc{binary\_c} \cite{Hurley2000}, which is potentially more relevant for low-mass stars. This yields $\sigma_{M_{\rm He, mod}} = 1\,M_{\odot}$ for stars with $M_{\rm ini}=12$--15~$M_{\odot}$, corresponding to the single-star equivalent initial mass for SN\,2018gj and typical of Type~II~SN progenitors.
Meanwhile, rotation can extend the MS lifetime by approximately 20\% \cite{2012Ekstrom} for stars with $M_{\rm ini} \gtrsim 7~M_{\odot}$; we therefore adopted $\sigma_{\log(t), \rm mod} = {\rm log}_{10}(1.2t)-{\rm log}_{10}(t) \approx 0.08$\,dex on a log scale. 

In the error covariance matrix (Eq.\,\eqref{eq:Cov}) we introduce a correlation coefficient $\rho$ between the model uncertainties of final He-core mass and age. 
The $\rho$ should be positive and relatively large, as both parameters are affected simultaneously by the same source of error (i.e. chemical mixing). For example, the impact of convective uncertainties on stellar structure and evolution was explored by varying relevant parameters \cite{2020Kaiser}.
The correlation coefficient of the resulting He-core mass and stellar lifetime (their Table~1) is 0.90 for the $M_{\rm ini} = 15~M_{\odot}$ star.
We note, however, that both quantities may also be affected by other physical processes, such as rotation, at the same time.
Besides, we assume no correlation between the model uncertainty of final H-envelope mass and that of any other parameter, since its uncertainty is dominated by another source of error (i.e. the mass-loss prescriptions), with negligible influences from variations in stellar structures.

With the above method we find the binary-progenitor probability, $P_{\rm BIN}$, should be very close to 1. While $P_{\rm BIN}$ increases as $\rho$ approaches 1, it remains insensitive to the value of $\rho$; specifically, it remains greater than 99\% even when $\rho$ is reduced to 0.2. In the unphysical scenario where $\rho$ becomes as low as 0.1 or 0.0, $P_{\rm BIN}$ only decreases to 88\% or 68\%, respectively. 
Moreover, if a slightly higher prior has been used (i.e. $P({\rm BIN})=P(\rm SIN)=0.5$, as in \cite{Bostroem2023}; essentially being agnostic to the expected occurrence of progenitors with binary history), the resulting posterior $P_{\rm BIN}$ increases to 91\% and 76\% for $\rho = 0.1$ and 0.0, further supporting the robustness of the binary origin.

Note that the stellar age distribution in the environment of SN\,2018gj has a non-negligible error, including that at the progenitor age expected from single-star evolution (i.e. log($t$/yr) = 7.1--7.3). We repeated the calculation by adopting $w_k$ from the last 1000 walkers to account for this uncertainty. In all cases, $P_{\rm BIN}$ remains almost unchanged. Therefore, our result robustly establishes SN\,2018gj as arising from an interacting binary progenitor.

The estimated lower limit of the final He-core mass of SN\,2018gj, $M_{\rm He}\approx3.2~M_\odot$, in case of high variability due to pulsations (Sec.~\ref{detection.sec}), would still be only barely consistent with the ages expected from single-star models assuming high mixing, such as in \textsc{posydon} (dashed line in Fig.\,\ref{fig:fig4}). 
Moreover, this lower mass estimate would be in tension with the progenitor mass independently inferred from nebular spectroscopy \cite{Teja2023}. 
Although the explosion phase within a pulsation cycle can change the decline rate of the plateau in optical light curves, it does not effectively affect the plateau duration and decay tail \cite{2025Bronner}. 
After accounting for radial pulsations as demonstrated in \cite{2025Bronner}, and applying the same scaling relation, we find that the resulting variation in the final H-envelope mass is negligible to the current observational uncertainties. Consequently, we argue that the evidence supporting a binary progenitor for SN~2018gj remains robust against the effects of large-amplitude pulsations.

\section{Modelling pre-supernova evolution}

In order to better understand the progenitor of SN\,2018gj, we perform a detailed modelling of binary systems that can give rise to such an event. In this investigation, we use the next-generation binary population synthesis framework, \textsc{posydon} (v2) \cite{posydonv2.ref} (expanding its previous public version \cite{posydon.ref}), which is based on extended grids of detailed binary systems with \textsc{mesa} \cite{mesa.ref1, 2013Paxton,2015Paxton, 2018Paxton, 2019Paxton}. These detailed binary models are crucial to follow accurately the response of both stars during multiple episodes of binary mass transfer, including the reverse mass transfer phases and eventual merger that are found to be important for this progenitor. \textsc{posydon} also provides self-consistent treatment of tidal evolution, contact phases, and rotation in its dense binary grids. 
For binaries with a non-degenerate accretor, initially all the material transferred by the donor through Roche-lobe overflow (RLOF) is accepted by the gainer and the accretion is restricted when the spun-up mass gainer reaches its critical rotation.  
For the common envelope (CE) evolution, we assume an efficiency of $\alpha_{\rm CE}=1$ in the $\alpha$-$\lambda$ formalism \cite{1984Webbink,1988Livio} and calculate the binding energy parameter $\lambda$ based on the detailed stellar profile of the donor star when the CE is triggered. All massive progenitors reach up to carbon core depletion, which is only decades away from core collapse. We exclude core-collapse progenitors that are expected to implode to black holes without a transient event according to \cite{2020Patton} mapping to the pre-SN state and \cite{Ertl+2016} explodability criterion.

In this framework\footnote{Exact \textsc{posydon} commit used in this work is \href{https://github.com/POSYDON-code/POSYDON/tree/Dimitris_WD_mergers}{891c5897}.}, we conduct a comprehensive search of the binary parameter space for models that reproduce the empirical values of the core mass, age, and envelope mass of the SN\,2018gj progenitor, within its observational errorbars as discussed in Sections~\ref{detection.sec} and \ref{age.sec}. We use a primary initial mass range of $M_1$~=~5--250~$M_{\odot}$, a flat initial secondary-to-primary mass ratio distribution from 0.05 to 0.99, and an initial period distribution following \cite{Sana2012} between 0.75 and 3500 days.  We assume solar metallicity for simplicity, as in all our analysis in this study (although see the Supplementary materials). 
We evolved $5 \times 10^4$ systems on the parameter grids, performing an initial-final linear interpolation for the computed final properties of the binary models. 

\begin{figure*}[htbp]
    \centering
    \includegraphics[width=0.9\textwidth]{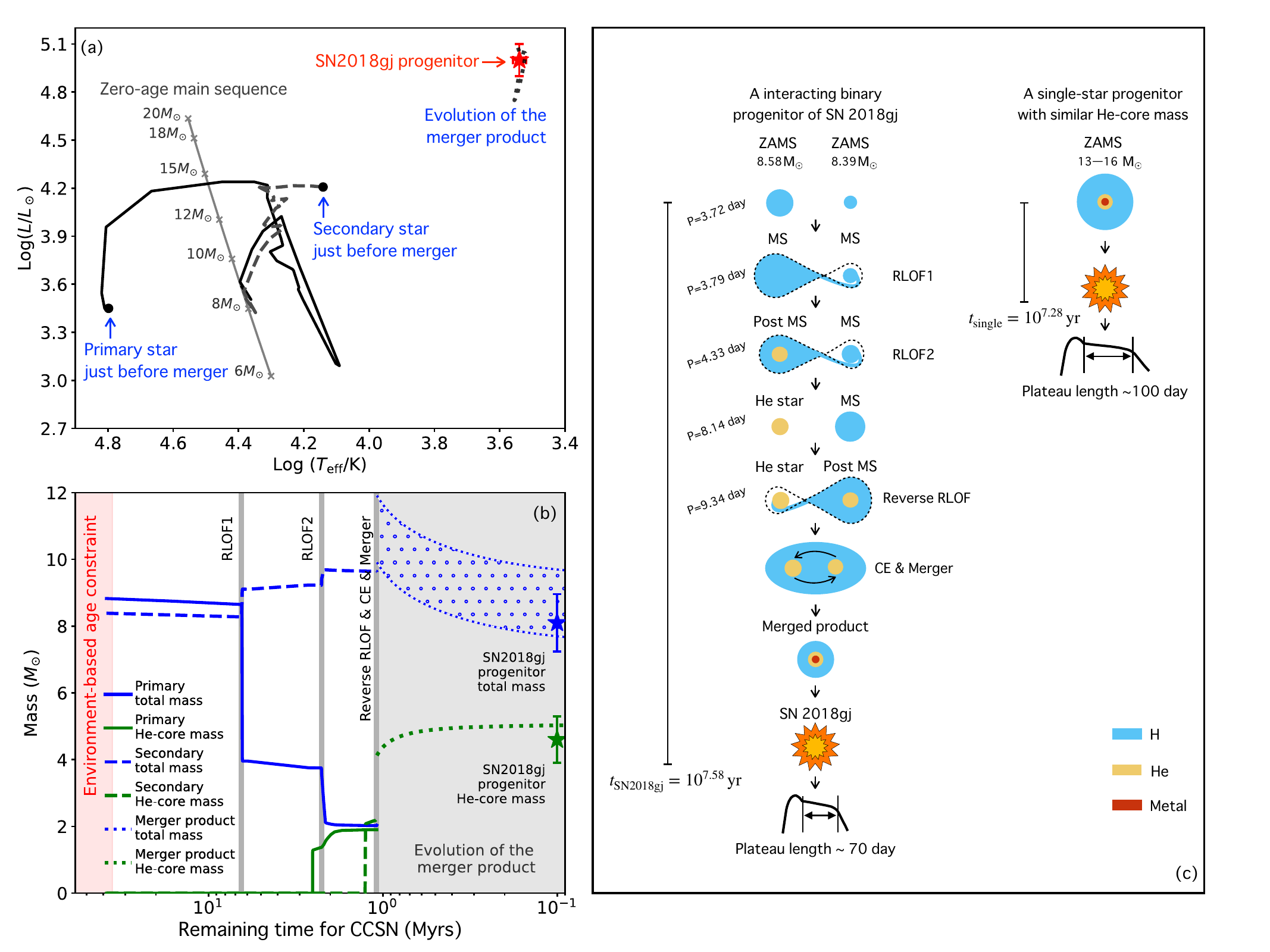}
    \caption{Pre-supernova evolution simulated with \textsc{posydon}.
    (a) Pre-SN evolution of the progenitor system of SN\,2018gj on the HRD.
    (b) Evolution of the total stellar mass and He-core mass of the primary star, secondary star, and merger product before explosion. Episodes of mass transfer are labeled in vertical, which are self-explanatory. The blue dotted region reflects the uncertainty of the total mass of the merger product with upper/lower limit corresponding to a conservative merging/partial CE ejection.
    Evolution continues up to core carbon depletion, expected to occur less than thousands of years before core collapse, with no significant further change in stellar mass or position on the HRD. (c) A schematic plot of the pre-SN evolution.}
    \label{fig:fig5}
\end{figure*}

Although binary scenarios including MS+MS mergers are found in principle to be able to produce a high mass core at longer delay-times, they have been excluded from our analysis as they are assumed to form a stellar structure similar to a more massive MS star, without being able to reproduce the uncommon core-to-envelope mass ratio. Similarly, scenarios involving the merging of an  evolved giant star with a MS companion, although allowed, are found not to reproduce the long inferred age of the SN event. Thus, we the find reverse merger scenario to be the main route towards a SN\,2018gj-like progenitor, plausibly reproducing all of its empirical values. 

%
The full evolution of an example reverse merger system, plausible for the evolution of SN~2018gj progenitor, is shown in Fig.\,\ref{fig:fig5}. The initial masses of the stellar components in the binary system ($\lesssim 9 M_{\odot}$) are the reason of the long-delay time until collapse. Although we consider the binary's evolution modeled with \textsc{mesa} binary systems until merging quite reliable, the merger product's subsequent evolution is more simplified (gray region in Fig.\,\ref{fig:fig5}b) but can still lead to physically-motivated conclusions. 
To more accurately estimate the eventual explodability of the merger product, we match the merger product with a single star model, whose He-core mass (if already formed during the evolution for both stars), CO-core mass, and mass-weighted average central He abundance are as similar as possible to the new cores of the merger product.
The merging of the two formed cores of the evolved stars enables a high core mass at such a long age. At the same time, as the mass loss during a merger involving two evolved stars is highly uncertain, we consider a range of post-merger envelope masses between two assumptions: on one hand the  conservative merger scenario (shown in the top of the star-hashed total mass of Fig.\,\ref{fig:fig5}b); on the other hand, the expected mass loss from unbinding of part of the donor's pre-CE envelope corresponding to the orbital energy released during spiraling-in at the CE phase \citep{Ivanova+2020}. A partial mass loss within this range can explain the low-mass envelope of SN~2018gj. Although in principle a non-conservative CE phase can form CSM around the merger product as seen in SN\,2018gj, we do not expect this to be present until explosion, after $\sim 1 $ Myr in our example system (although the CSM may be the indirect outcome of a different evolution of a merger product until explosion, compared to single stars).

The initial parameter space of binary systems that undergo a reverse merger and evolve towards a SN\,2018gj-like progenitor is shown in Fig.\,S2 (online). Such systems have a relatively low initial primary mass of $M_1$ = 8.3--9.3~$M_\odot$, an almost equal initial mass ratio of $q$ = 0.85--0.97, and a short initial period of $P$ = 3.0--6.5~days.  Binaries of slightly lower initial mass ratios do not lead to unstable reverse mass transfer, whereas wider systems start their first mass transfer not during but after the MS phase of the primary, resulting in SN progenitors inconsistent with that of SN\,2018gj.

Binary channels involving these kind of reverse mergers have been suggested to give rise to SN progenitors with long delay times. If the initial mass of the primary star is below the minimum mass threshold of $\sim8\,M_\odot$ for core-collapse SNe, these merger SN progenitors can exceed the lifetimes of all SNe originating from single-stars \cite{Zapartas2017, Zapartas2019}. There has been observational evidence of SN remnants \cite{Zapartas2017,Auchettl+2018} and compact object remnants \citep{Murphy+2024} that their inferred age favor scenarios of prior mass exchange or merging in a binary system, potentially reverse merging. Additionally, there are indications, although inconclusive, of prolonged SN ages \cite{Castrillo+2020}. The scenario discussed here for SN\,2018gj can be seen as a slightly higher mass analog of the current state and expected outcome of the $\phi$ Persei system \cite{Schootemeijer+2018}. 
Similar scenarios of reverse mergers involving two already-formed cores can lead to merger products with high angular momentum budget within their cores, suggested as potential channel for long gamma-ray bursts \citep{Fryer+2005,Tout+2011}.

\section{Conclusion}

SNe\,II-P are the most common type of core-collapse SNe in the local Universe and play critical roles in many aspects of astrophysics. Since decades ago theorists have predicted that they may originate not only from single stars but also from interacting binaries; however, robust observational evidence still remains elusive for this speculated binary progenitor channel.

In this paper, we report the first identification of a SN\,II-P progenitor with compelling evidence of binary origin in the case of SN 2018gj. The progenitor is detected on the pre-explosion HST images and has disappeared after explosion. The progenitor has an effective temperature of log($T_{\rm eff}$/K) = 3.54 $\pm$~0.01, a luminosity of log$(L/L_{\odot})$ = 5.0 $\pm$~0.1 and a final He-core mass of $M_{\rm He}$ = 4.6\,$\pm$\,0.7\,$M_\odot$.

While the directly detected progenitor exhibits deceptively typical temperature and luminosity of other SN\,II-P progenitors, it is located in an unexpectedly old environment, with age of log($t$/yr)$\ge7.6\pm0.05$, and displays an abnormally short plateau in the light curve, which corresponds to a very low final H-envelope mass of $M_{\rm H}=3.5 \pm 0.5\,M_{\odot}$. Both phenomena are inconsistent with expectations from single-star evolution, and we show that SN\,2018gj is unambiguously associated with a binary progenitor with a high probability of $>99\%$.

With state-of-the-art binary simulation code \textsc{posydon} we further reveal the initial parameters and the pre-SN evolution of the progenitor system of SN\,2018gj. The system is found to be initially a close binary consisting of two nearly equal-mass stars with primary mass $M_1$ = 8.3--9.3\,$M_\odot$, secondary-to-primary mass ratio $q$ = 0.85--0.97 and period $P$ = 3.0--6.5\,days. The system experienced a reverse merger scenario, which involves a complicated history of mass transfer back and forth between the two stars followed by a merger. The merger product corresponds to the direct progenitor of SN\,2018gj that we have detected on the pre-explosion image, and this scenario explains all the observational characteristics.

This study provides the first robust observational evidence for the long-sought interacting binary progenitor channel toward SNe\,II-P, which is key to decipher the origin of this major class of SNe as well as their observational diversity. Our study also establishes a successful methodology that can motivate future investigations to uncover a previously hidden population of SNe\,II-P from binaries in the wealth of data of existing and upcoming surveys such as CSST, Euclid, Roman, ZTF and LSST.

~\\

\begin{acknowledgments}
This work is supported by the Strategic Priority Research Program of the Chinese Academy of Sciences (XDB0550300), the National Natural Science Foundation of China (12303039, 12303051, 12261141690, and 12588202).
EZ and DS acknowledge support from the Hellenic Foundation for Research and Innovation (H.F.R.I.) under the ``3rd Call for H.F.R.I. Research Projects to support Post-Doctoral Researchers” (7933). EZ acknowledges useful discussion with Azalee Bostroem, Jared Goldberg and Natalia Ivanova. 
JJA acknowledges support provided through a grant (JWST-AR-04369.001-A) from the STScI under NASA contract NAS5-03127. 
MMB was supported by the Boninchi Foundation, the Swiss National Science Foundation (CRSII5\_21349), and the Swiss Government Excellence Scholarship. 
MF is supported by a Royal Society - Science Foundation Ireland University Research Fellowship.
MK was supported by the Swiss National Science Foundation Professorship grant (PI Fragos, PP00P2 176868). 
SG, CL, PMS and ET were supported by the Gordon and Betty Moore Foundation (PI Kalogera, GBMF8477 and GBMF12341). 
ZWL work is supported by the National Natural Science Foundation of China (12288102), the Strategic Priority Research Program of the Chinese Academy of Sciences (XDB1160303), the International Centre of Supernovae (ICESUN), Yunnan Key Laboratory of Supernova Research (202302AN360001) and the Yunnan Revitalization Talent Support Program ``YunLing Scholar" project.
YY’s research is partially supported by the Tsinghua University Dushi Program.
JFL acknowledges support from the New Cornerstone Science Foundation through the New Cornerstone Investigator Program and the XPLORER PRIZE.

This research is based on observations made with the NASA/ESA Hubble Space Telescope obtained from the Space Telescope Science Institute, which is operated by the Association of Universities for Research in Astronomy, Inc., under NASA contract NAS 5–26555. 
This work made use of v2.2 of the Binary Population and Spectral Synthesis (BPASS) models as described in \cite{bpass.ref} and \cite{2018Stanway}.
\end{acknowledgments}

\begin{widetext}

\appendix

\section{Distance, reddening and metallicity}\label{sec:metallicity}

SN\,2018gj occurred in the nearby galaxy NGC\,6217 with a heliocentric redshift of $z$ = 0.004540 (from \cite{redshift.ref} and the NASA/IPAC Extragalactic Database). This corresponds to a cosmic-expansion velocity of 1774\,$\pm$\,18\,km\,s$^{-1}$, which has been corrected for the influence of the Virgo cluster, Great Attracter and Shapley supercluster \cite{virgo.ref}. Adopting a Hubble constant ($H_0$) of 73.30\,$\pm$\,1.04\,km\,s$^{-1}$\,Mpc$^{-1}$ \cite{H0.ref}, the distance to NGC\,6217 was estimated to be 24.2\,$\pm$\,0.4\,Mpc and this value was used throughout this paper.

SN\,2018gj has a total line-of-sight reddening of $E(B-V) = 0.08 \pm 0.02$\,mag, consisting of a Galactic reddening of $E(B-V) = 0.04$\,mag from the IRSA-Galactic Dust Reddening and Extinction map \cite{sfd.ref} and a host-galaxy reddening of $E(B-V) = 0.04\,\pm\,0.02$\,mag as estimated with the Na\,{\sc i}\,D absorption feature in the SN spectra \cite{Teja2023}. Given the very low value of reddening, variation of the extinction law has a very small influence on our results and we used a standard Galactic extinction law with $R_V = 3.1$ \cite{avlaw.ref} in this work.

NGC\,6217 is a nearly face-on, Seyfert\,2 disk galaxy with a total mass of about 10$^{10.8}M_{\odot}$ and a high metallicity of 12\,+\,log(O/H) = 8.98 near the center \cite{1991vanDriel, 1994Storchi-Bergmann}. The metallicities for disk galaxies decrease outward and, for massive galaxies of $>$10$^{10.5}M_{\odot}$, the metallicity gradients are very similar to each other \citep{2023Lu}. Therefore, we used the metallicity gradient of $-0.020\,\pm\,0.006$\,dex\,kpc$^{-1}$ of M81, which is also a massive Seyfert\,2 galaxy and resembles NGC\,6217 very much in the morphology, as an estimate of that for NGC\,6217 \citep{2012Patterson}. 
This corresponds to a solar metallicity at the position of SN\,2018gj. In addition, Ref.\,\cite{Teja2023} found that the evolution of Fe\,{\sc ii}\,$\lambda$\,5169\,\AA\ pseudo-equivalent width of SN\,2018gj is consistent with hydrodynamical models with solar or half-solar metallicities. 
Combining these estimates, we adopted a solar metallicity for SN\,2018gj in the whole analysis. It is worth mentioning that the conclusion of this paper will not be affected even if SN\,2018gj has a sub-solar metallicity, where if anything, lower wind mass-loss rates \cite{Vink+2001,Antoniadis+2024} and less efficient Roche-lobe stripping \cite{2017gotberg,2024Hovis} are expected.

\section{Fitting for the star formation history}\label{sec:SFH}
In order to derive the stellar age distribution of the SN environment, we fit stars inside the large circular region, $R_{\rm env}$, with model stellar populations with a hierarchical Bayesian approach \cite{Maund2016, Sun2021}. The model populations were from the \textsc{bpass} v2.2 population and spectral synthesis \cite{bpass.ref} including interacting binaries, adopting a Salpeter initial mass function with a power-law slope of $\alpha$\,=\,$-$2.35 \cite{imf.ref} and a maximum stellar mass of 300\,$M_\odot$. For mono-age populations of log($t$/yr)\,=\,6.0, 6.1, 6.2, ... , \textsc{bpass} provides number distributions on CMDs \cite{h_f_stevance_2022_7340797}. 

We note that \textsc{bpass} has only calculated synthetic colors and magnitudes in some selected filter systems (i.e. Johnson-Cousins and HST/WFPC2 filters); those for the HST/ACS and HST/WFC3 filters are not provided. Therefore, we transformed the stellar magnitudes from the observed filters to their closest filters in \textsc{bpass} (i.e. from WFC3/F555W to WFPC2/F555W). To find the transformation relation, we used \textsc{pysynphot} to calculate the synthetic magnitudes of blackbodies with different temperatures and fit the magnitude difference between the two similar filters as a 3rd-order polynomial of color. The magnitude difference is very small (typically $\lesssim$0.1~mag for most stars) compared with the photometric uncertainties. Distance, reddening and metallicity were the same as SN\,2018gj's. In particular, it is reasonable to assume a uniform extinction across this area since the SN environment is very simple and without recent star formation.

For the $i$-th star, its likelihood of belonging to the $k$-th population was calculated by comparing its color and magnitude with model predictions on a series of CMDs
\begin{equation}
p^{i,k} = \prod \limits_{j}p^{i,k}_j,
\end{equation}
and
\begin{equation}
\begin{aligned}
p^{i,k}_j = & \iint \frac{1}{\sqrt{2\pi} \sigma^i_{c, j}} \exp\left[-\frac{1}{2}\left(\frac{c^i_j - c_j^{\rm mod}}{\sigma^i_{c, j}}\right)^2\right] 
\times \frac{1}{\sqrt{2\pi} \sigma^i_{m, j}} \exp\left[-\frac{1}{2}\left(\frac{m^i_j - m_j^{\rm mod}}{\sigma^i_{m, j}}\right)^2\right] \\ &
\times p^k_j(c_j^{\rm mod}, m_j^{\rm mod})
dc_j^{\rm mod}dm_j^{\rm mod},
\label{likelihood}
\end{aligned}
\end{equation}
where, in the $j$-th CMD, $m_j^{\rm mod}$, $m^i_j$ and $\sigma^i_{m, j}$ are the model magnitude, observed magnitude and its uncertainty; $c_j^{\rm mod}$, $c^i_j$ and $\sigma^i_{c, j}$ are those for the color; and $p^k_j(c_j^{\rm mod}, m_j^{\rm mod})$ is the model probability density distribution. In this work, we used two CMDs of (F555W$-$F625W, F625W) and (F625W$-$F814W, F814W), i.e. $c_1 = m_{\rm F555W}-m_{\rm F625W}$, $c_2 = m_{\rm F625W}-m_{\rm F814W}$, $m_1 =  m_{\rm F625W}$, and $m_2 = m_{\rm F814W}$. If a star was not detected in a filter, we could only provide a detection limit for its color and/or magnitude; in this case, we substituted the Gaussian function in \eqref{likelihood} with an error function. For example, if in the $j$-th CMD the $i$-th star has a magnitude measurement but only a lower limit for its color, $c^i_j \geq c^{\rm lim}_j \pm \sigma^{\rm lim}_{c, j}$, then $p^{i,k}_j$ takes the form
\begin{equation}
\begin{aligned}
p^{i,k}_j = & \iint \frac{1}{2}\left[1+ {\rm erf}\left(\frac{c_j^{\rm mod} - c^{\rm lim}_j}{\sqrt{2}\sigma^{\rm lim}_{c, j}}\right) \right] 
\times \frac{1}{\sqrt{2\pi} \sigma^i_{m, j}}\exp\left[-\frac{1}{2}\left(\frac{m^i_j - m_j^{\rm mod}}{\sigma^i_{m, j}}\right)^2\right] \\ &
\times p^k_j(c_j^{\rm mod}, m_j^{\rm mod})
dc_j^{\rm mod}dm_j^{\rm mod}.
\end{aligned}
\end{equation}

The observed stars can be considered as arising from a mixture of mono-age populations, each having a relative weight of $w_k$, i.e. the fraction of observed stars belonging to the $k$-th population. The weights then provide information of the stellar age distribution in the SN environment. In a hierarchical Bayesian framework, their posterior probability distributions can be expressed as
\begin{equation}
p (w_1, w_2, w_3, ... | {\rm data} ) = \prod \limits_{i}^{N_{\rm star}}\left(\sum\limits_{k}^{N_{\rm pop}}w_k p^{i,k} \right),
\end{equation}
where we have adopted a flat prior for the weights of different mono-age populations. We used a Markov-Chain Monte-Carlo (MCMC) method to solve for the posterior probability distributions.

\section{Hydrodynamical modeling for bolometric light curves}\label{sec:bollcmodl}

The multi-band photometric data of SN\,2018gj were taken from Ref.\,\cite{Teja2023}, and the UV–optical–IR (pseudo-)bolometric light curve was reconstructed using the \textsc{SuperBol} package \cite{2018Nicholl}. The blackbody-corrected light curve was adopted for our analysis. As shown in Fig.\,\ref{fig:bol_lc}, we present the bolometric light curve together with the model predictions from \textsc{mesa} and \textsc{stella}, following a similar approach to Ref.\,\cite{Teja2023} but with the initial mass updated to 15~$M_{\odot}$.

\section{Initial parameters of the binary system}\label{sec:preNN}

From our modeled systems, we find that approximately 0.1\% of all successful core–collapse SNe in our population would match the observational properties of SN\,2018gj within uncertainties based on a Gaussian treatment of the observational constraints on the luminosity, effective temperature, final He-core mass, final H-envelope mass, and age of SN\,2018gj's progenitor.
Their initial parameters distributions are shown in Fig.\,\ref{fig:NN}.
This suggests that SN\,2018gj–like explosions form a small but non-negligible subset of the broader short–plateau Type~II–P population, which itself represents roughly \(\sim4\text{--}5\%\) of core–collapse SNe in theoretical studies (e.g., \cite{Eldridge2018,2019Eldridge,Hiramatsu2021}) and a few percent observationally (e.g., \cite{2014Anderson}). A full comparison with the complete short–plateau SN population is outside the scope of this work.

\begin{figure}
    \centering
    \includegraphics[width=0.5\textwidth]{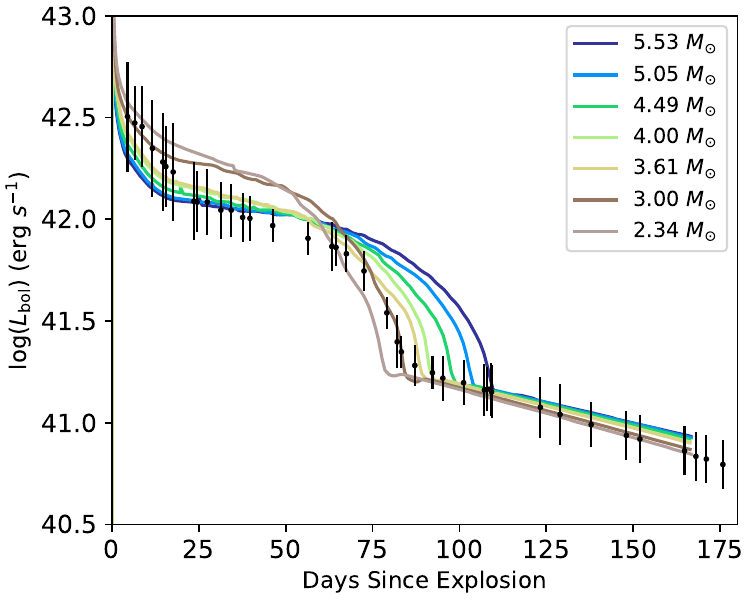}
    \caption{The UV-optical-IR pseudo-bolometric light curve of SN\,2018gj. The observations (black points) are compared with  models (color‑coded by final H‑envelope mass) for which the wind mass‑loss rate scaling factor was varied from 2.25 to 3.70 in steps of 0.25.}
    \label{fig:bol_lc}
\end{figure}

\begin{figure}
    \centering
    \includegraphics[width=0.8\textwidth]{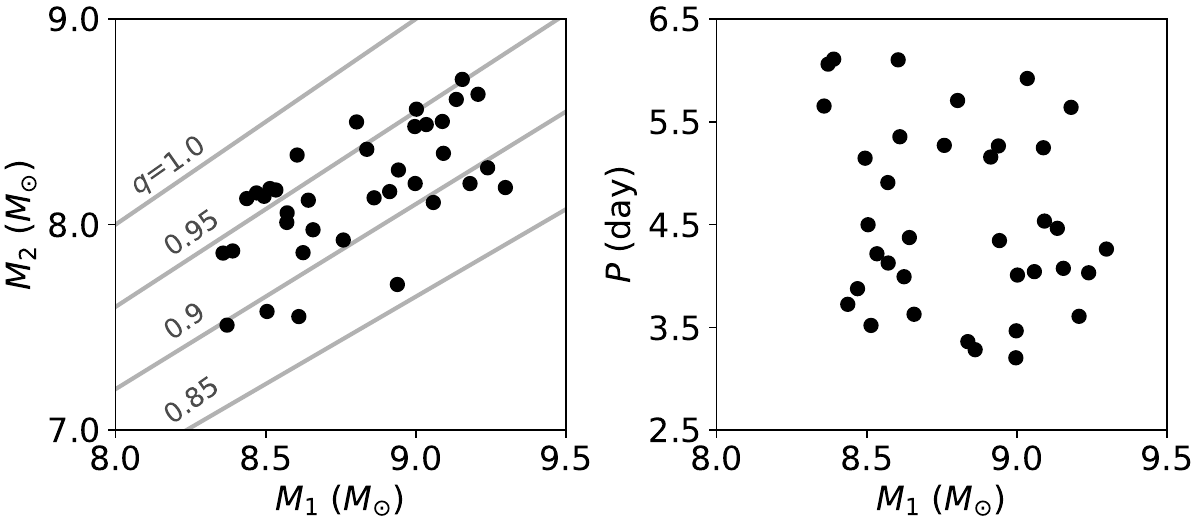}
    \caption{Initial primary mass ($M_1$), secondary mass ($M_2$) and orbital period for possible reverse-merger models that match the observed constraints for SN\,2018gj's progenitor. For reference, the gray lines of constant secondary-to-primary mass ratios ($q$) are plotted in the left panel.}
    \label{fig:NN}
\end{figure}

\end{widetext}  

\bibliographystyle{scibull}
\bibliography{SNbib_scibull}

@ARTICLE{2024Utrobin,
       author = {{Utrobin}, V.~P. and {Chugai}, N.~N.},
        title = "{Revisiting short-plateau SN 2018gj}",
      journal = {\apss},
         year = 2024,
        month = may,
       volume = {369},
       number = {5},
          eid = {49},
        pages = {49},
          doi = {10.1007/s10509-024-04311-9},
archivePrefix = {arXiv},
       eprint = {2405.12867},
 primaryClass = {astro-ph.HE},
       adsurl = {https://ui.adsabs.harvard.edu/abs/2024Ap&SS.369...49U}
}

@ARTICLE{Antoniadis+2024,
       author = {{Antoniadis}, K. and {Bonanos}, A.~Z. and {de Wit}, S. and {Zapartas}, E. and {Munoz-Sanchez}, G. and {Maravelias}, G.},
        title = "{Establishing a mass-loss rate relation for red supergiants in the Large Magellanic Cloud}",
      journal = {\aap},
         year = 2024,
        month = jun,
       volume = {686},
          eid = {A88},
        pages = {A88},
          doi = {10.1051/0004-6361/202449383},
archivePrefix = {arXiv},
       eprint = {2401.15163},
 primaryClass = {astro-ph.SR},
       adsurl = {https://ui.adsabs.harvard.edu/abs/2024A&A...686A..88A}
}

@ARTICLE{Niu2024,
       author = {{Niu}, Zexi and {Sun}, Ning-Chen and {Liu}, Jifeng},
        title = "{Discovery of a dusty yellow supergiant progenitor for the Type IIb SN 2017gkk}",
      journal = {\apjl},
         year = 2024,
        month = jul,
       volume = {970},
       number = {1},
          eid = {L9},
        pages = {L9},
          doi = {10.3847/2041-8213/ad5f20},
archivePrefix = {arXiv},
       eprint = {2407.03721},
 primaryClass = {astro-ph.HE},
       adsurl = {https://ui.adsabs.harvard.edu/abs/2024ApJ...970L...9N}
}

@ARTICLE{Zhao2025,
       author = {{Zhao}, Yi-Han and {Sun}, Ning-Chen and {Wu}, Junjie and {Niu}, Zexi and {Hong}, Xinyi and {Huang}, Yinhan and {Maund}, Justyn R. and {Xi}, Qiang and {Xiang}, Danfeng and {Liu}, Jifeng},
        title = "{Exclusion of a direct progenitor detection for the Type Ic SN 2017ein based on late-time observations}",
      journal = {\apjl},
         year = 2025,
        month = feb,
       volume = {980},
       number = {1},
          eid = {L6},
        pages = {L6},
          doi = {10.3847/2041-8213/adad5d},
archivePrefix = {arXiv},
       eprint = {2411.17969},
 primaryClass = {astro-ph.HE},
       adsurl = {https://ui.adsabs.harvard.edu/abs/2025ApJ...980L...6Z}
}

@ARTICLE{sun2020,
       author = {{Sun}, Ning-Chen and {Maund}, Jusytn R. and {Hirai}, Ryosuke and {Crowther}, Paul A. and {Podsiadlowski}, Philipp},
        title = "{Origins of Type Ibn SNe 2006jc/2015G in interacting binaries and implications for pre-SN eruptions}",
      journal = {\mnras},
         year = 2020,
        month = feb,
       volume = {491},
       number = {4},
        pages = {6000-6019},
          doi = {10.1093/mnras/stz3431},
archivePrefix = {arXiv},
       eprint = {1909.07999},
 primaryClass = {astro-ph.SR},
       adsurl = {https://ui.adsabs.harvard.edu/abs/2020MNRAS.491.6000S}
}

@ARTICLE{1991vanDriel,
       author = {{van Driel}, W. and {Buta}, R.~J.},
        title = "{A study of the ringed galaxies NGC 2273, 4826 and 6217.}",
      journal = {\aap},
         year = 1991,
        month = may,
       volume = {245},
        pages = {7},
       adsurl = {https://ui.adsabs.harvard.edu/abs/1991A&A...245....7V}
}

@ARTICLE{1994Storchi-Bergmann,
       author = {{Storchi-Bergmann}, Thaisa and {Calzetti}, Daniela and {Kinney}, Anne L.},
        title = "{Ultraviolet to near-infrared spectral distributions of star-forming galaxies: metallicity and age effects}",
      journal = {\apj},
         year = 1994,
        month = jul,
       volume = {429},
        pages = {572},
          doi = {10.1086/174345},
       adsurl = {https://ui.adsabs.harvard.edu/abs/1994ApJ...429..572S}
}

@ARTICLE{2018Nicholl,
       author = {{Nicholl}, Matt},
        title = "{SuperBol: a user-friendly Python routine for bolometric light curves}",
      journal = {Research Notes of the American Astronomical Society},
         year = 2018,
        month = dec,
       volume = {2},
       number = {4},
          eid = {230},
        pages = {230},
          doi = {10.3847/2515-5172/aaf799},
       adsurl = {https://ui.adsabs.harvard.edu/abs/2018RNAAS...2..230N}
}

@ARTICLE{stella.ref1,
       author = {{Blinnikov}, S. and {Sorokina}, E.},
        title = "{Type Ia supernova models: latest developments}",
      journal = {\apss},
         year = 2004,
        month = feb,
       volume = {290},
       number = {1},
        pages = {13-28},
          doi = {10.1023/B:ASTR.0000022161.03559.42},
archivePrefix = {arXiv},
       eprint = {astro-ph/0212530},
 primaryClass = {astro-ph},
       adsurl = {https://ui.adsabs.harvard.edu/abs/2004Ap&SS.290...13B}
}

@ARTICLE{mesa.ref1,
       author = {{Paxton}, Bill and {Bildsten}, Lars and {Dotter}, Aaron and {Herwig}, Falk and {Lesaffre}, Pierre and {Timmes}, Frank},
        title = "{Modules for experiments in stellar astrophysics (MESA)}",
      journal = {\apjs},
         year = 2011,
        month = jan,
       volume = {192},
       number = {1},
          eid = {3},
        pages = {3},
          doi = {10.1088/0067-0049/192/1/3},
archivePrefix = {arXiv},
       eprint = {1009.1622},
 primaryClass = {astro-ph.SR},
       adsurl = {https://ui.adsabs.harvard.edu/abs/2011ApJS..192....3P}
}

@ARTICLE{2016Schneider,
       author = {{Schneider}, F.~R.~N. and {Podsiadlowski}, Ph. and {Langer}, N. and {Castro}, N. and {Fossati}, L.},
        title = "{Rejuvenation of stellar mergers and the origin of magnetic fields in massive stars}",
      journal = {\mnras},
         year = 2016,
        month = apr,
       volume = {457},
       number = {3},
        pages = {2355-2365},
          doi = {10.1093/mnras/stw148},
archivePrefix = {arXiv},
       eprint = {1601.05084},
 primaryClass = {astro-ph.SR},
       adsurl = {https://ui.adsabs.harvard.edu/abs/2016MNRAS.457.2355S}
}

@misc{2025Laplace,
  author  = {{Laplace}, Eva and {Bronner}, Vincent A. and {Schneider}, Fabian R.~N. and {Podsiadlowski}, Philipp},
  title   = "{Pulsations change the structures of massive stars before they explode: interpreting the nearby supernova SN 2023ixf}",
  howpublished = {arXiv: 2508.11088},
  year    = {2025}
}

@ARTICLE{2025Bronner,
       author = {{Bronner}, V.~A. and {Laplace}, E. and {Schneider}, F.~R.~N. and {Podsiadlowski}, Ph.},
        title = "{Explosions of pulsating red supergiants: A natural pathway for the diversity of Type II-P/L supernovae}",
      journal = {\aap},
         year = 2025,
        month = nov,
       volume = {703},
          eid = {A61},
        pages = {A61},
          doi = {10.1051/0004-6361/202554642},
archivePrefix = {arXiv},
       eprint = {2508.11077},
 primaryClass = {astro-ph.SR},
       adsurl = {https://ui.adsabs.harvard.edu/abs/2025A&A...703A..61B}
}

@ARTICLE{Bostroem2023,
       author = {{Bostroem}, K. Azalee and {Zapartas}, Emmanouil and {Koplitz}, Brad and {Williams}, Benjamin F. and {Tran}, Debby and {Dolphin}, Andrew},
        title = "{Considering the single and binary origins of the Type IIP SN 2017eaw}",
      journal = {\aj},
         year = 2023,
        month = dec,
       volume = {166},
       number = {6},
          eid = {255},
        pages = {255},
          doi = {10.3847/1538-3881/acffc7},
archivePrefix = {arXiv},
       eprint = {2310.01498},
 primaryClass = {astro-ph.HE},
       adsurl = {https://ui.adsabs.harvard.edu/abs/2023AJ....166..255B}
}

@ARTICLE{Podsiadlowski1992,
       author = {{Podsiadlowski}, Ph. and {Joss}, P.~C. and {Hsu}, J.~J.~L.},
        title = "{Presupernova evolution in massive interacting binaries}",
      journal = {\apj},
         year = 1992,
        month = may,
       volume = {391},
        pages = {246},
          doi = {10.1086/171341},
       adsurl = {https://ui.adsabs.harvard.edu/abs/1992ApJ...391..246P}
}

@ARTICLE{2024Zimmerman,
       author = {{Zimmerman}, E.~A. and {Irani}, I. and {Chen}, P. and {Gal-Yam}, A. and {Schulze}, S. and {Perley}, D.~A. and {Sollerman}, J. and {Filippenko}, A.~V. and {Shenar}, T. and {Yaron}, O. and {Shahaf}, S. and {Bruch}, R.~J. and {Ofek}, E.~O. and {De Cia}, A. and {Brink}, T.~G. and {Yang}, Y. and {Vasylyev}, S.~S. and {Ben Ami}, S. and {Aubert}, M. and {Badash}, A. and {Bloom}, J.~S. and {Brown}, P.~J. and {De}, K. and {Dimitriadis}, G. and {Fransson}, C. and {Fremling}, C. and {Hinds}, K. and {Horesh}, A. and {Johansson}, J.~P. and {Kasliwal}, M.~M. and {Kulkarni}, S.~R. and {Kushnir}, D. and {Martin}, C. and {Matuzewski}, M. and {McGurk}, R.~C. and {Miller}, A.~A. and {Morag}, J. and {Neil}, J.~D. and {Nugent}, P.~E. and {Post}, R.~S. and {Prusinski}, N.~Z. and {Qin}, Y. and {Raichoor}, A. and {Riddle}, R. and {Rowe}, M. and {Rusholme}, B. and {Sfaradi}, I. and {Sjoberg}, K.~M. and {Soumagnac}, M. and {Stein}, R.~D. and {Strotjohann}, N.~L. and {Terwel}, J.~H. and {Wasserman}, T. and {Wise}, J. and {Wold}, A. and {Yan}, L. and {Zhang}, K.},
        title = "{The complex circumstellar environment of supernova 2023ixf}",
      journal = {\nat},
         year = 2024,
        month = mar,
       volume = {627},
       number = {8005},
        pages = {759-762},
          doi = {10.1038/s41586-024-07116-6},
archivePrefix = {arXiv},
       eprint = {2310.10727},
 primaryClass = {astro-ph.HE},
       adsurl = {https://ui.adsabs.harvard.edu/abs/2024Natur.627..759Z}
}

@ARTICLE{2011Brott,
       author = {{Brott}, I. and {de Mink}, S.~E. and {Cantiello}, M. and {Langer}, N. and {de Koter}, A. and {Evans}, C.~J. and {Hunter}, I. and {Trundle}, C. and {Vink}, J.~S.},
        title = "{Rotating massive main-sequence stars. I. Grids of evolutionary models and isochrones}",
      journal = {\aap},
         year = 2011,
        month = jun,
       volume = {530},
          eid = {A115},
        pages = {A115},
          doi = {10.1051/0004-6361/201016113},
archivePrefix = {arXiv},
       eprint = {1102.0530},
 primaryClass = {astro-ph.SR},
       adsurl = {https://ui.adsabs.harvard.edu/abs/2011A&A...530A.115B}
}

@ARTICLE{2012Ekstrom,
       author = {{Ekstr{\"o}m}, S. and {Georgy}, C. and {Eggenberger}, P. and {Meynet}, G. and {Mowlavi}, N. and {Wyttenbach}, A. and {Granada}, A. and {Decressin}, T. and {Hirschi}, R. and {Frischknecht}, U. and {Charbonnel}, C. and {Maeder}, A.},
        title = "{Grids of stellar models with rotation. I. Models from 0.8 to 120 M$_{{\ensuremath{\odot}}}$ at solar metallicity ($Z$ = 0.014)}",
      journal = {\aap},
         year = 2012,
        month = jan,
       volume = {537},
          eid = {A146},
        pages = {A146},
          doi = {10.1051/0004-6361/201117751},
archivePrefix = {arXiv},
       eprint = {1110.5049},
 primaryClass = {astro-ph.SR},
       adsurl = {https://ui.adsabs.harvard.edu/abs/2012A&A...537A.146E}
}

@ARTICLE{2020Kaiser,
       author = {{Kaiser}, Etienne A. and {Hirschi}, Raphael and {Arnett}, W. David and {Georgy}, Cyril and {Scott}, Laura J.~A. and {Cristini}, Andrea},
        title = "{Relative importance of convective uncertainties in massive stars}",
      journal = {\mnras},
         year = 2020,
        month = aug,
       volume = {496},
       number = {2},
        pages = {1967-1989},
          doi = {10.1093/mnras/staa1595},
archivePrefix = {arXiv},
       eprint = {2006.01877},
 primaryClass = {astro-ph.SR},
       adsurl = {https://ui.adsabs.harvard.edu/abs/2020MNRAS.496.1967K}
}

@ARTICLE{2014Sukhbold,
       author = {{Sukhbold}, Tuguldur and {Woosley}, S.~E.},
        title = "{The compactness of presupernova stellar cores}",
      journal = {\apj},
         year = 2014,
        month = mar,
       volume = {783},
       number = {1},
          eid = {10},
        pages = {10},
          doi = {10.1088/0004-637X/783/1/10},
archivePrefix = {arXiv},
       eprint = {1311.6546},
 primaryClass = {astro-ph.SR},
       adsurl = {https://ui.adsabs.harvard.edu/abs/2014ApJ...783...10S}
}

@ARTICLE{2025Zapartas_1,
       author = {{Zapartas}, E. and {Fox}, O.~D. and {Su}, J. and {Souropanis}, D. and {Drout}, M.~R. and {Rocha}, K.~A. and {van Dyk}, S.~D. and {Williams}, B.~F. and {Briel}, M. and {Renzo}, M. and {Andrews}, J.~J. and {Fragos}, T. and {Gossage}, S. and {Kruckow}, M.~U. and {Liotine}, C. and {Ryder}, S.~D. and {Srivastava}, P.~M. and {Teng}, E.},
        title = "{The demographics of binary companions to stripped-envelope supernovae: confronting population synthesis models with observations}",
      journal = {\mnras},
         year = 2025,
        month = dec,
          doi = {10.1093/mnras/staf2208},
archivePrefix = {arXiv},
       eprint = {2508.12677},
 primaryClass = {astro-ph.SR},
       adsurl = {https://ui.adsabs.harvard.edu/abs/2025MNRAS.tmp.2097Z}
}

@ARTICLE{2018Morozova,
       author = {{Morozova}, Viktoriya and {Piro}, Anthony L. and {Valenti}, Stefano},
        title = "{Measuring the progenitor masses and dense circumstellar material of Type II supernovae}",
      journal = {\apj},
         year = 2018,
        month = may,
       volume = {858},
       number = {1},
          eid = {15},
        pages = {15},
          doi = {10.3847/1538-4357/aab9a6},
archivePrefix = {arXiv},
       eprint = {1709.04928},
 primaryClass = {astro-ph.HE},
       adsurl = {https://ui.adsabs.harvard.edu/abs/2018ApJ...858...15M}
}

@ARTICLE{2020Beasor,
       author = {{Beasor}, Emma R. and {Davies}, Ben and {Smith}, Nathan and {van Loon}, Jacco Th and {Gehrz}, Robert D. and {Figer}, Donald F.},
        title = "{A new mass-loss rate prescription for red supergiants}",
      journal = {\mnras},
         year = 2020,
        month = mar,
       volume = {492},
       number = {4},
        pages = {5994-6006},
          doi = {10.1093/mnras/staa255},
archivePrefix = {arXiv},
       eprint = {2001.07222},
 primaryClass = {astro-ph.SR},
       adsurl = {https://ui.adsabs.harvard.edu/abs/2020MNRAS.492.5994B}
}

@ARTICLE{2019Dessart,
       author = {{Dessart}, Luc and {Hillier}, D. John},
        title = "{The difficulty of inferring progenitor masses from type-II-Plateau supernova light curves}",
      journal = {\aap},
         year = 2019,
        month = may,
       volume = {625},
          eid = {A9},
        pages = {A9},
          doi = {10.1051/0004-6361/201834732},
archivePrefix = {arXiv},
       eprint = {1903.04840},
 primaryClass = {astro-ph.SR},
       adsurl = {https://ui.adsabs.harvard.edu/abs/2019A&A...625A...9D}
}

@ARTICLE{2025Kilpatrick,
       author = {{Kilpatrick}, Charles D. and {Suresh}, Aswin and {Davis}, Kyle W. and {Drout}, Maria R. and {Foley}, Ryan J. and {Gagliano}, Alexander and {Jacobson-Gal{\'a}n}, Wynn V. and {Kaur}, Ravjit and {Taggart}, Kirsty and {Vazquez}, Jason},
        title = "{The Type II SN 2025pht in NGC 1637: A Red Supergiant with Carbon-rich Circumstellar Dust as the First JWST Detection of a Supernova Progenitor Star}",
      journal = {\apjl},
         year = 2025,
        month = oct,
       volume = {992},
       number = {1},
          eid = {L10},
        pages = {L10},
          doi = {10.3847/2041-8213/ae04de},
archivePrefix = {arXiv},
       eprint = {2508.10994},
 primaryClass = {astro-ph.HE},
       adsurl = {https://ui.adsabs.harvard.edu/abs/2025ApJ...992L..10K}
}

@ARTICLE{2012Patterson,
       author = {{Patterson}, Maria T. and {Walterbos}, Rene A.~M. and {Kennicutt}, Robert C. and {Chiappini}, Cristina and {Thilker}, David A.},
        title = "{An oxygen abundance gradient into the outer disc of M81}",
      journal = {\mnras},
         year = 2012,
        month = may,
       volume = {422},
       number = {1},
        pages = {401-419},
          doi = {10.1111/j.1365-2966.2012.20616.x},
archivePrefix = {arXiv},
       eprint = {1202.0308},
 primaryClass = {astro-ph.CO},
       adsurl = {https://ui.adsabs.harvard.edu/abs/2012MNRAS.422..401P}
}

@ARTICLE{2025Fangql,
       author = {{Fang}, Qiliang and {Maeda}, Keiichi and {Ye}, Haonan and {Moriya}, Takashi J. and {Matsumoto}, Tatsuya},
        title = "{Diversity in hydrogen-rich envelope mass of Type II supernovae. I. Plateau phase light-curve modeling}",
      journal = {\apj},
         year = 2025,
        month = jan,
       volume = {978},
       number = {1},
          eid = {35},
        pages = {35},
          doi = {10.3847/1538-4357/ad8b19},
archivePrefix = {arXiv},
       eprint = {2404.01776},
 primaryClass = {astro-ph.HE},
       adsurl = {https://ui.adsabs.harvard.edu/abs/2025ApJ...978...35F}
}

@software{curtis_mccully_2018_1482019,
  author       = {Curtis McCully and
                  Steve Crawford and
                  Gabor Kovacs and
                  Erik Tollerud and
                  Edward Betts and
                  Larry Bradley and
                  Matt Craig and
                  James Turner and
                  Ole Streicher and
                  Brigitta Sipocz and
                  Thomas Robitaille and
                  Christoph Deil},
  title        = {astropy/astroscrappy: v1.0.5 Zenodo release},
  month        = nov,
  year         = 2018,
  publisher    = {Zenodo},
  version      = {v1.0.5},
  doi          = {10.5281/zenodo.1482019},
  url          = {https://doi.org/10.5281/zenodo.1482019},
}

@ARTICLE{LACosmic.ref,
       author = {{van Dokkum}, Pieter G.},
        title = "{Cosmic-ray rejection by Laplacian edge detection}",
      journal = {\pasp},
         year = 2001,
        month = nov,
       volume = {113},
       number = {789},
        pages = {1420-1427},
          doi = {10.1086/323894},
archivePrefix = {arXiv},
       eprint = {astro-ph/0108003},
 primaryClass = {astro-ph},
       adsurl = {https://ui.adsabs.harvard.edu/abs/2001PASP..113.1420V}
}

@ARTICLE{2020Sun,
       author = {{Sun}, Ning-Chen and {Maund}, Justyn R. and {Crowther}, Paul A.},
        title = "{The changing-type SN 2014C may come from an 11-M$_{{\ensuremath{\odot}}}$ star stripped by binary interaction and violent eruption}",
      journal = {\mnras},
         year = 2020,
        month = oct,
       volume = {497},
       number = {4},
        pages = {5118-5135},
          doi = {10.1093/mnras/staa2277},
archivePrefix = {arXiv},
       eprint = {2003.09325},
 primaryClass = {astro-ph.SR},
       adsurl = {https://ui.adsabs.harvard.edu/abs/2020MNRAS.497.5118S}
}

@ARTICLE{2015VanDyk,
       author = {{Van Dyk}, Schuyler D. and {Lee}, Janice C. and {Anderson}, Jay and {Andrews}, Jennifer E. and {Calzetti}, Daniela and {Bright}, Stacey N. and {Ubeda}, Leonardo and {Smith}, Linda J. and {Sabbi}, Elena and {Grebel}, Eva K. and {Herrero}, Artemio and {de Mink}, Selma E.},
        title = "{LEGUS discovery of a light echo around supernova 2012aw}",
      journal = {\apj},
         year = 2015,
        month = jun,
       volume = {806},
       number = {2},
          eid = {195},
        pages = {195},
          doi = {10.1088/0004-637X/806/2/195},
archivePrefix = {arXiv},
       eprint = {1504.08323},
 primaryClass = {astro-ph.SR},
       adsurl = {https://ui.adsabs.harvard.edu/abs/2015ApJ...806..195V}
}

@ARTICLE{Sun2023a,
       author = {{Sun}, Ning-Chen and {Maund}, Justyn R. and {Crowther}, Paul A.},
        title = "{A UV census of the environments of stripped-envelope supernovae}",
      journal = {\mnras},
         year = 2023,
        month = may,
       volume = {521},
       number = {2},
        pages = {2860-2873},
          doi = {10.1093/mnras/stad690},
archivePrefix = {arXiv},
       eprint = {2209.05283},
 primaryClass = {astro-ph.SR},
       adsurl = {https://ui.adsabs.harvard.edu/abs/2023MNRAS.521.2860S}
}

@ARTICLE{Maund2016,
       author = {{Maund}, Justyn R. and {Ramirez-Ruiz}, Enrico},
        title = "{A high mass progenitor for the Type Ic supernova 2007gr inferred from its environment}",
      journal = {\mnras},
         year = 2016,
        month = mar,
       volume = {456},
       number = {3},
        pages = {3175-3185},
          doi = {10.1093/mnras/stv2760},
archivePrefix = {arXiv},
       eprint = {1609.07740},
 primaryClass = {astro-ph.SR},
       adsurl = {https://ui.adsabs.harvard.edu/abs/2016MNRAS.456.3175M}
}

@ARTICLE{2019VanDyk,
       author = {{Van Dyk}, Schuyler D. and {Zheng}, WeiKang and {Maund}, Justyn R. and {Brink}, Thomas G. and {Srinivasan}, Sundar and {Andrews}, Jennifer E. and {Smith}, Nathan and {Leonard}, Douglas C. and {Morozova}, Viktoriya and {Filippenko}, Alexei V. and {Conner}, Brody and {Milisavljevic}, Dan and {de Jaeger}, Thomas and {Long}, Knox S. and {Isaacson}, Howard and {Crossfield}, Ian J.~M. and {Kosiarek}, Molly R. and {Howard}, Andrew W. and {Fox}, Ori D. and {Kelly}, Patrick L. and {Piro}, Anthony L. and {Littlefair}, Stuart P. and {Dhillon}, Vik S. and {Wilson}, Richard and {Butterley}, Timothy and {Yunus}, Sameen and {Channa}, Sanyum and {Jeffers}, Benjamin T. and {Falcon}, Edward and {Ross}, Timothy W. and {Hestenes}, Julia C. and {Stegman}, Samantha M. and {Zhang}, Keto and {Kumar}, Sahana},
        title = "{The Type II-plateau supernova 2017eaw in NGC 6946 and its red supergiant progenitor}",
      journal = {\apj},
         year = 2019,
        month = apr,
       volume = {875},
       number = {2},
          eid = {136},
        pages = {136},
          doi = {10.3847/1538-4357/ab1136},
archivePrefix = {arXiv},
       eprint = {1903.03872},
 primaryClass = {astro-ph.HE},
       adsurl = {https://ui.adsabs.harvard.edu/abs/2019ApJ...875..136V}
}

@ARTICLE{Hurley2000,
       author = {{Hurley}, Jarrod R. and {Pols}, Onno R. and {Tout}, Christopher A.},
        title = "{Comprehensive analytic formulae for stellar evolution as a function of mass and metallicity}",
      journal = {\mnras},
         year = 2000,
        month = jul,
       volume = {315},
       number = {3},
        pages = {543-569},
          doi = {10.1046/j.1365-8711.2000.03426.x},
archivePrefix = {arXiv},
       eprint = {astro-ph/0001295},
 primaryClass = {astro-ph},
       adsurl = {https://ui.adsabs.harvard.edu/abs/2000MNRAS.315..543H}
}

@ARTICLE{2018Boubert,
       author = {{Boubert}, D. and {Guillochon}, J. and {Hawkins}, K. and {Ginsburg}, I. and {Evans}, N.~W. and {Strader}, J.},
        title = "{Revisiting hypervelocity stars after Gaia DR2}",
      journal = {\mnras},
         year = 2018,
        month = sep,
       volume = {479},
       number = {2},
        pages = {2789-2795},
          doi = {10.1093/mnras/sty1601},
archivePrefix = {arXiv},
       eprint = {1804.10179},
 primaryClass = {astro-ph.GA},
       adsurl = {https://ui.adsabs.harvard.edu/abs/2018MNRAS.479.2789B}
}

@ARTICLE{2022Marchetti,
       author = {{Marchetti}, Tommaso and {Evans}, Fraser A. and {Rossi}, Elena Maria},
        title = "{Gaia DR3 in 6D: the search for fast hypervelocity stars and constraints on the galactic centre environment}",
      journal = {\mnras},
         year = 2022,
        month = sep,
       volume = {515},
       number = {1},
        pages = {767-774},
          doi = {10.1093/mnras/stac1777},
archivePrefix = {arXiv},
       eprint = {2206.06962},
 primaryClass = {astro-ph.GA},
       adsurl = {https://ui.adsabs.harvard.edu/abs/2022MNRAS.515..767M}
}

@ARTICLE{Izzard2004,
       author = {{Izzard}, Robert G. and {Tout}, Christopher A. and {Karakas}, Amanda I. and {Pols}, Onno R.},
        title = "{A new synthetic model for asymptotic giant branch stars}",
      journal = {\mnras},
         year = 2004,
        month = may,
       volume = {350},
       number = {2},
        pages = {407-426},
          doi = {10.1111/j.1365-2966.2004.07446.x},
archivePrefix = {arXiv},
       eprint = {astro-ph/0402403},
 primaryClass = {astro-ph},
       adsurl = {https://ui.adsabs.harvard.edu/abs/2004MNRAS.350..407I}
}

@ARTICLE{1988Livio,
       author = {{Livio}, Mario and {Soker}, Noam},
        title = "{The common envelope phase in the evolution of binary stars}",
      journal = {\apj},
         year = 1988,
        month = jun,
       volume = {329},
        pages = {764},
          doi = {10.1086/166419},
       adsurl = {https://ui.adsabs.harvard.edu/abs/1988ApJ...329..764L}
}

@ARTICLE{2019Eldridge,
       author = {{Eldridge}, J.~J. and {Guo}, N.-Y. and {Rodrigues}, N. and {Stanway}, E.~R. and {Xiao}, L.},
        title = "{Supernova lightCURVE POPulation synthesis II: validation against supernovae with an observed progenitor}",
      journal = {\pasa},
         year = 2019,
        month = nov,
       volume = {36},
          eid = {e041},
        pages = {e041},
          doi = {10.1017/pasa.2019.31},
archivePrefix = {arXiv},
       eprint = {1908.07762},
 primaryClass = {astro-ph.SR},
       adsurl = {https://ui.adsabs.harvard.edu/abs/2019PASA...36...41E}
}

@ARTICLE{2013Eldridge,
       author = {{Eldridge}, John J. and {Fraser}, Morgan and {Smartt}, Stephen J. and {Maund}, Justyn R. and {Crockett}, R. Mark},
        title = "{The death of massive stars - II. Observational constraints on the progenitors of Type Ibc supernovae}",
      journal = {\mnras},
         year = 2013,
        month = nov,
       volume = {436},
       number = {1},
        pages = {774-795},
          doi = {10.1093/mnras/stt1612},
archivePrefix = {arXiv},
       eprint = {1301.1975},
 primaryClass = {astro-ph.SR},
       adsurl = {https://ui.adsabs.harvard.edu/abs/2013MNRAS.436..774E}
}

@ARTICLE{1984Webbink,
       author = {{Webbink}, R.~F.},
        title = "{Double white dwarfs as progenitors of R Coronae Borealis stars and type I supernovae.}",
      journal = {\apj},
         year = 1984,
        month = feb,
       volume = {277},
        pages = {355-360},
          doi = {10.1086/161701},
       adsurl = {https://ui.adsabs.harvard.edu/abs/1984ApJ...277..355W}
}

@ARTICLE{2020Schneider,
       author = {{Schneider}, F.~R.~N. and {Ohlmann}, S.~T. and {Podsiadlowski}, Ph and {R{\"o}pke}, F.~K. and {Balbus}, S.~A. and {Pakmor}, R.},
        title = "{Long-term evolution of a magnetic massive merger product}",
      journal = {\mnras},
         year = 2020,
        month = jul,
       volume = {495},
       number = {3},
        pages = {2796-2812},
          doi = {10.1093/mnras/staa1326},
archivePrefix = {arXiv},
       eprint = {2005.05335},
 primaryClass = {astro-ph.SR},
       adsurl = {https://ui.adsabs.harvard.edu/abs/2020MNRAS.495.2796S}
}

@ARTICLE{2020Patton,
       author = {{Patton}, Rachel A. and {Sukhbold}, Tuguldur},
        title = "{Towards a realistic explosion landscape for binary population synthesis}",
      journal = {\mnras},
         year = 2020,
        month = dec,
       volume = {499},
       number = {2},
        pages = {2803-2816},
          doi = {10.1093/mnras/staa3029},
archivePrefix = {arXiv},
       eprint = {2005.03055},
 primaryClass = {astro-ph.SR},
       adsurl = {https://ui.adsabs.harvard.edu/abs/2020MNRAS.499.2803P}
}

@ARTICLE{posydonv2.ref,
       author = {{Andrews}, Jeff J. and {Bavera}, Simone S. and {Briel}, Max and {Chattaraj}, Abhishek and {Dotter}, Aaron and {Fragos}, Tassos and {Gallegos-Garcia}, Monica and {Gossage}, Seth and {Kalogera}, Vicky and {Kasdagli}, Eirini and {Katsaggelos}, Aggelos and {Kimball}, Chase and {Kovlakas}, Konstantinos and {Kruckow}, Matthias U. and {Liotine}, Camille and {Misra}, Devina and {Rocha}, Kyle A. and {Souropanis}, Dimitris and {Srivastava}, Philipp M. and {Sun}, Meng and {Teng}, Elizabeth and {Xing}, Zepei and {Zapartas}, Emmanouil and {Zevin}, Michael},
        title = "{POSYDON Version 2: Population Synthesis with Detailed Binary-evolution Simulations across a Cosmological Range of Metallicities}",
      journal = {\apjs},
         year = 2025,
        month = nov,
       volume = {281},
       number = {1},
          eid = {3},
        pages = {3},
          doi = {10.3847/1538-4365/adfb78},
archivePrefix = {arXiv},
       eprint = {2411.02376},
 primaryClass = {astro-ph.GA},
       adsurl = {https://ui.adsabs.harvard.edu/abs/2025ApJS..281....3A}
}

@ARTICLE{posydon.ref,
       author = {{Fragos}, Tassos and {Andrews}, Jeff J. and {Bavera}, Simone S. and {Berry}, Christopher P.~L. and {Coughlin}, Scott and {Dotter}, Aaron and {Giri}, Prabin and {Kalogera}, Vicky and {Katsaggelos}, Aggelos and {Kovlakas}, Konstantinos and {Lalvani}, Shamal and {Misra}, Devina and {Srivastava}, Philipp M. and {Qin}, Ying and {Rocha}, Kyle A. and {Rom{\'a}n-Garza}, Jaime and {Serra}, Juan Gabriel and {Stahle}, Petter and {Sun}, Meng and {Teng}, Xu and {Trajcevski}, Goce and {Tran}, Nam Hai and {Xing}, Zepei and {Zapartas}, Emmanouil and {Zevin}, Michael},
        title = "{POSYDON: a general-purpose population synthesis code with detailed binary-evolution simulations}",
      journal = {\apjs},
         year = 2023,
        month = feb,
       volume = {264},
       number = {2},
          eid = {45},
        pages = {45},
          doi = {10.3847/1538-4365/ac90c1},
archivePrefix = {arXiv},
       eprint = {2202.05892},
 primaryClass = {astro-ph.SR},
       adsurl = {https://ui.adsabs.harvard.edu/abs/2023ApJS..264...45F}
}

@ARTICLE{2018Stanway,
       author = {{Stanway}, E.~R. and {Eldridge}, J.~J.},
        title = "{Re-evaluating old stellar populations}",
      journal = {\mnras},
         year = 2018,
        month = sep,
       volume = {479},
       number = {1},
        pages = {75-93},
          doi = {10.1093/mnras/sty1353},
archivePrefix = {arXiv},
       eprint = {1805.08784},
 primaryClass = {astro-ph.GA},
       adsurl = {https://ui.adsabs.harvard.edu/abs/2018MNRAS.479...75S}
}

@ARTICLE{Eldridge2018,
       author = {{Eldridge}, J.~J. and {Xiao}, L. and {Stanway}, E.~R. and {Rodrigues}, N. and {Guo}, N. -Y.},
        title = "{Supernova lightCURVE POPulation synthesis I: including interacting binaries is key to understanding the diversity of type II supernova lightcurves}",
      journal = {\pasa},
         year = 2018,
        month = dec,
       volume = {35},
          eid = {e049},
        pages = {e049},
          doi = {10.1017/pasa.2018.47},
archivePrefix = {arXiv},
       eprint = {1811.00282},
 primaryClass = {astro-ph.SR},
       adsurl = {https://ui.adsabs.harvard.edu/abs/2018PASA...35...49E}
}

@ARTICLE{Hicken2017,
       author = {{Hicken}, Malcolm and {Friedman}, Andrew S. and {Blondin}, Stephane and {Challis}, Peter and {Berlind}, Perry and {Calkins}, Mike and {Esquerdo}, Gil and {Matheson}, Thomas and {Modjaz}, Maryam and {Rest}, Armin and {Kirshner}, Robert P.},
        title = "{Type II supernova light curves and spectra from the CfA}",
      journal = {\apjs},
         year = 2017,
        month = nov,
       volume = {233},
       number = {1},
          eid = {6},
        pages = {6},
          doi = {10.3847/1538-4365/aa8ef4},
archivePrefix = {arXiv},
       eprint = {1706.01030},
 primaryClass = {astro-ph.HE},
       adsurl = {https://ui.adsabs.harvard.edu/abs/2017ApJS..233....6H}
}

@ARTICLE{Valenti2016,
       author = {{Valenti}, S. and {Howell}, D.~A. and {Stritzinger}, M.~D. and {Graham}, M.~L. and {Hosseinzadeh}, G. and {Arcavi}, I. and {Bildsten}, L. and {Jerkstrand}, A. and {McCully}, C. and {Pastorello}, A. and {Piro}, A.~L. and {Sand}, D. and {Smartt}, S.~J. and {Terreran}, G. and {Baltay}, C. and {Benetti}, S. and {Brown}, P. and {Filippenko}, A.~V. and {Fraser}, M. and {Rabinowitz}, D. and {Sullivan}, M. and {Yuan}, F.},
        title = "{The diversity of Type II supernova versus the similarity in their progenitors}",
      journal = {\mnras},
         year = 2016,
        month = jul,
       volume = {459},
       number = {4},
        pages = {3939-3962},
          doi = {10.1093/mnras/stw870},
archivePrefix = {arXiv},
       eprint = {1603.08953},
 primaryClass = {astro-ph.SR},
       adsurl = {https://ui.adsabs.harvard.edu/abs/2016MNRAS.459.3939V}
}

@ARTICLE{dolphot.ref,
       author = {{Dolphin}, Andrew E.},
        title = "{WFPC2 stellar photometry with HSTPHOT}",
      journal = {\pasp},
         year = 2000,
        month = oct,
       volume = {112},
       number = {776},
        pages = {1383-1396},
          doi = {10.1086/316630},
archivePrefix = {arXiv},
       eprint = {astro-ph/0006217},
 primaryClass = {astro-ph},
       adsurl = {https://ui.adsabs.harvard.edu/abs/2000PASP..112.1383D}
}

@ARTICLE{Sun2022,
       author = {{Sun}, Ning-Chen and {Maund}, Justyn R. and {Crowther}, Paul A. and {Hirai}, Ryosuke and {Kashapov}, Amir and {Liu}, Ji-Feng and {Liu}, Liang-Duan and {Zapartas}, Emmanouil},
        title = "{An environmental analysis of the Type Ib SN 2019yvr and the possible presence of an inflated binary companion}",
      journal = {\mnras},
         year = 2022,
        month = mar,
       volume = {510},
       number = {3},
        pages = {3701-3715},
          doi = {10.1093/mnras/stab3768},
archivePrefix = {arXiv},
       eprint = {2111.06471},
 primaryClass = {astro-ph.SR},
       adsurl = {https://ui.adsabs.harvard.edu/abs/2022MNRAS.510.3701S}
}

@ARTICLE{2025Zapartas,
       author = {{Zapartas}, E. and {de Wit}, S. and {Antoniadis}, K. and {Mu{\~n}oz-Sanchez}, G. and {Souropanis}, D. and {Bonanos}, A.~Z. and {Maravelias}, G. and {Kovlakas}, K. and {Kruckow}, M.~U. and {Fragos}, T. and {Andrews}, J.~J. and {Bavera}, S.~S. and {Briel}, M. and {Gossage}, S. and {Kasdagli}, E. and {Rocha}, K.~A. and {Sun}, M. and {Srivastava}, P.~M. and {Xing}, Z.},
        title = "{The effect of mass loss in models of red supergiants in the Small Magellanic Cloud}",
      journal = {\aap},
         year = 2025,
        month = may,
       volume = {697},
          eid = {A167},
        pages = {A167},
          doi = {10.1051/0004-6361/202452401},
archivePrefix = {arXiv},
       eprint = {2410.07335},
 primaryClass = {astro-ph.SR},
       adsurl = {https://ui.adsabs.harvard.edu/abs/2025A&A...697A.167Z}
}

@ARTICLE{2025Fang,
       author = {{Fang}, Qiliang and {Moriya}, Takashi J. and {Maeda}, Keiichi},
        title = "{Red supergiant problem viewed from the nebular phase spectroscopy of type II supernovae}",
      journal = {\apj},
         year = 2025,
        month = jun,
       volume = {986},
       number = {1},
          eid = {39},
        pages = {39},
          doi = {10.3847/1538-4357/adceae},
archivePrefix = {arXiv},
       eprint = {2504.14502},
 primaryClass = {astro-ph.HE},
       adsurl = {https://ui.adsabs.harvard.edu/abs/2025ApJ...986...39F}
}

@ARTICLE{2025Wagg,
       author = {{Wagg}, Tom and {Dalcanton}, Julianne J. and {Renzo}, Mathieu and {Breivik}, Katelyn and {Orr}, Matthew E. and {Price-Whelan}, Adrian M. and {Cruz}, Akaxia and {Brooks}, Alyson and {Steinwandel}, Ulrich P. and {Bellm}, Eric C.},
        title = "{Delayed and Displaced: The Impact of Binary Interactions on Core-collapse SN Feedback}",
      journal = {\aj},
         year = 2025,
        month = sep,
       volume = {170},
       number = {3},
          eid = {192},
        pages = {192},
          doi = {10.3847/1538-3881/adf55e},
archivePrefix = {arXiv},
       eprint = {2504.17903},
 primaryClass = {astro-ph.SR},
       adsurl = {https://ui.adsabs.harvard.edu/abs/2025AJ....170..192W}
}

@ARTICLE{2020Goldberg,
       author = {{Goldberg}, Jared A. and {Bildsten}, Lars},
        title = "{The value of progenitor radius measurements for explosion modeling of Type II-Plateau supernovae}",
      journal = {\apjl},
         year = 2020,
        month = jun,
       volume = {895},
       number = {2},
          eid = {L45},
        pages = {L45},
          doi = {10.3847/2041-8213/ab9300},
archivePrefix = {arXiv},
       eprint = {2005.07290},
 primaryClass = {astro-ph.SR},
       adsurl = {https://ui.adsabs.harvard.edu/abs/2020ApJ...895L..45G}
}

@ARTICLE{Schneider2024,
       author = {{Schneider}, F.~R.~N. and {Podsiadlowski}, Ph. and {Laplace}, E.},
        title = "{Pre-supernova evolution and final fate of stellar mergers and accretors of binary mass transfer}",
      journal = {\aap},
         year = 2024,
        month = jun,
       volume = {686},
          eid = {A45},
        pages = {A45},
          doi = {10.1051/0004-6361/202347854},
archivePrefix = {arXiv},
       eprint = {2403.03984},
 primaryClass = {astro-ph.SR},
       adsurl = {https://ui.adsabs.harvard.edu/abs/2024A&A...686A..45S}
}

@dataset{h_f_stevance_2022_7340797,
  author       = {H.F. Stevance},
  title        = {BPASSv2.2.1 starter kit},
  month        = nov,
  year         = 2022,
  publisher    = {Zenodo},
  doi          = {10.5281/zenodo.7340797},
  url          = {https://doi.org/10.5281/zenodo.7340797}
}

@ARTICLE{2019Paxton,
       author = {{Paxton}, Bill and {Smolec}, R. and {Schwab}, Josiah and {Gautschy}, A. and {Bildsten}, Lars and {Cantiello}, Matteo and {Dotter}, Aaron and {Farmer}, R. and {Goldberg}, Jared A. and {Jermyn}, Adam S. and {Kanbur}, S.~M. and {Marchant}, Pablo and {Thoul}, Anne and {Townsend}, Richard H.~D. and {Wolf}, William M. and {Zhang}, Michael and {Timmes}, F.~X.},
        title = "{Modules for Experiments in Stellar Astrophysics (MESA): pulsating variable stars, rotation, convective boundaries, and energy conservation}",
      journal = {\apjs},
         year = 2019,
        month = jul,
       volume = {243},
       number = {1},
          eid = {10},
        pages = {10},
          doi = {10.3847/1538-4365/ab2241},
archivePrefix = {arXiv},
       eprint = {1903.01426},
 primaryClass = {astro-ph.SR},
       adsurl = {https://ui.adsabs.harvard.edu/abs/2019ApJS..243...10P}
}

@ARTICLE{2018Paxton,
       author = {{Paxton}, Bill and {Schwab}, Josiah and {Bauer}, Evan B. and {Bildsten}, Lars and {Blinnikov}, Sergei and {Duffell}, Paul and {Farmer}, R. and {Goldberg}, Jared A. and {Marchant}, Pablo and {Sorokina}, Elena and {Thoul}, Anne and {Townsend}, Richard H.~D. and {Timmes}, F.~X.},
        title = "{Modules for Experiments in Stellar Astrophysics (MESA): convective boundaries, element diffusion, and massive star explosions}",
      journal = {\apjs},
         year = 2018,
        month = feb,
       volume = {234},
       number = {2},
          eid = {34},
        pages = {34},
          doi = {10.3847/1538-4365/aaa5a8},
archivePrefix = {arXiv},
       eprint = {1710.08424},
 primaryClass = {astro-ph.SR},
       adsurl = {https://ui.adsabs.harvard.edu/abs/2018ApJS..234...34P}
}

@ARTICLE{2015Paxton,
       author = {{Paxton}, Bill and {Marchant}, Pablo and {Schwab}, Josiah and {Bauer}, Evan B. and {Bildsten}, Lars and {Cantiello}, Matteo and {Dessart}, Luc and {Farmer}, R. and {Hu}, H. and {Langer}, N. and {Townsend}, R.~H.~D. and {Townsley}, Dean M. and {Timmes}, F.~X.},
        title = "{Modules for Experiments in Stellar Astrophysics (MESA): binaries, pulsations, and explosions}",
      journal = {\apjs},
         year = 2015,
        month = sep,
       volume = {220},
       number = {1},
          eid = {15},
        pages = {15},
          doi = {10.1088/0067-0049/220/1/15},
archivePrefix = {arXiv},
       eprint = {1506.03146},
 primaryClass = {astro-ph.SR},
       adsurl = {https://ui.adsabs.harvard.edu/abs/2015ApJS..220...15P}
}

@ARTICLE{2013Paxton,
       author = {{Paxton}, Bill and {Cantiello}, Matteo and {Arras}, Phil and {Bildsten}, Lars and {Brown}, Edward F. and {Dotter}, Aaron and {Mankovich}, Christopher and {Montgomery}, M.~H. and {Stello}, Dennis and {Timmes}, F.~X. and {Townsend}, Richard},
        title = "{Modules for Experiments in Stellar Astrophysics (MESA): planets, oscillations, rotation, and massive stars}",
      journal = {\apjs},
         year = 2013,
        month = sep,
       volume = {208},
       number = {1},
          eid = {4},
        pages = {4},
          doi = {10.1088/0067-0049/208/1/4},
archivePrefix = {arXiv},
       eprint = {1301.0319},
 primaryClass = {astro-ph.SR},
       adsurl = {https://ui.adsabs.harvard.edu/abs/2013ApJS..208....4P}
}

@ARTICLE{imf.ref,
       author = {{Salpeter}, Edwin E.},
        title = "{The luminosity function and stellar evolution.}",
      journal = {\apj},
         year = 1955,
        month = jan,
       volume = {121},
        pages = {161},
          doi = {10.1086/145971},
       adsurl = {https://ui.adsabs.harvard.edu/abs/1955ApJ...121..161S}
}

@ARTICLE{2004Maund,
       author = {{Maund}, Justyn R. and {Smartt}, Stephen J. and {Kudritzki}, Rolf P. and {Podsiadlowski}, Philipp and {Gilmore}, Gerard F.},
        title = "{The massive binary companion star to the progenitor of supernova 1993J}",
      journal = {\nat},
         year = 2004,
        month = jan,
       volume = {427},
       number = {6970},
        pages = {129-131},
          doi = {10.1038/nature02161},
archivePrefix = {arXiv},
       eprint = {astro-ph/0401090},
 primaryClass = {astro-ph},
       adsurl = {https://ui.adsabs.harvard.edu/abs/2004Natur.427..129M}
}

@ARTICLE{bpass.ref,
       author = {{Eldridge}, J.~J. and {Stanway}, E.~R. and {Xiao}, L. and {McClelland}, L.~A.~S. and {Taylor}, G. and {Ng}, M. and {Greis}, S.~M.~L. and {Bray}, J.~C.},
        title = "{Binary Population and Spectral Synthesis Version 2.1: construction, observational verification, and new results}",
      journal = {\pasa},
         year = 2017,
        month = nov,
       volume = {34},
          eid = {e058},
        pages = {e058},
          doi = {10.1017/pasa.2017.51},
archivePrefix = {arXiv},
       eprint = {1710.02154},
 primaryClass = {astro-ph.SR},
       adsurl = {https://ui.adsabs.harvard.edu/abs/2017PASA...34...58E}
}

@ARTICLE{marcs.ref,
       author = {{Gustafsson}, B. and {Edvardsson}, B. and {Eriksson}, K. and {J{\o}rgensen}, U.~G. and {Nordlund}, {\r{A}}. and {Plez}, B.},
        title = "{A grid of MARCS model atmospheres for late-type stars. I. Methods and general properties}",
      journal = {\aap},
         year = 2008,
        month = aug,
       volume = {486},
       number = {3},
        pages = {951-970},
          doi = {10.1051/0004-6361:200809724},
archivePrefix = {arXiv},
       eprint = {0805.0554},
 primaryClass = {astro-ph},
       adsurl = {https://ui.adsabs.harvard.edu/abs/2008A&A...486..951G}
}

@ARTICLE{sfd.ref,
       author = {{Schlafly}, Edward F. and {Finkbeiner}, Douglas P.},
        title = "{Measuring reddening with Sloan Digital Sky Survey stellar spectra and recalibrating SFD}",
      journal = {\apj},
         year = 2011,
        month = aug,
       volume = {737},
       number = {2},
          eid = {103},
        pages = {103},
          doi = {10.1088/0004-637X/737/2/103},
archivePrefix = {arXiv},
       eprint = {1012.4804},
 primaryClass = {astro-ph.GA},
       adsurl = {https://ui.adsabs.harvard.edu/abs/2011ApJ...737..103S}
}

@ARTICLE{Xiang2024,
       author = {{Xiang}, Danfeng and {Mo}, Jun and {Wang}, Xiaofeng and {Wang}, Lingzhi and {Zhang}, Jujia and {Lin}, Han and {Chen}, Liyang and {Song}, Cuiying and {Liu}, Liang-Duan and {Wang}, Zhenyu and {Li}, Gaici},
        title = "{The red supergiant progenitor of Type II supernova 2024ggi}",
      journal = {\apjl},
         year = 2024,
        month = jul,
       volume = {969},
       number = {1},
          eid = {L15},
        pages = {L15},
          doi = {10.3847/2041-8213/ad54b3},
archivePrefix = {arXiv},
       eprint = {2405.07699},
 primaryClass = {astro-ph.HE},
       adsurl = {https://ui.adsabs.harvard.edu/abs/2024ApJ...969L..15X}
}

@ARTICLE{Rui2019,
       author = {{Rui}, Liming and {Wang}, Xiaofeng and {Mo}, Jun and {Xiang}, Danfeng and {Zhang}, Jujia and {Maund}, Justyn R. and {Gal-Yam}, Avishy and {Wang}, Lifan and {Zhang}, Tianmeng},
        title = "{Probing the final-stage progenitor evolution for Type IIP supernova 2017eaw in NGC 6946}",
      journal = {\mnras},
         year = 2019,
        month = may,
       volume = {485},
       number = {2},
        pages = {1990-2000},
          doi = {10.1093/mnras/stz503},
archivePrefix = {arXiv},
       eprint = {1902.06181},
 primaryClass = {astro-ph.HE},
       adsurl = {https://ui.adsabs.harvard.edu/abs/2019MNRAS.485.1990R}
}

@ARTICLE{2023Lu,
       author = {{Lu}, Shengdong and {Zhu}, Kai and {Cappellari}, Michele and {Li}, Ran and {Mao}, Shude and {Xu}, Dandan},
        title = "{MaNGA DynPop - II. Global stellar population, gradients, and star-formation histories from integral-field spectroscopy of 10K galaxies: link with galaxy rotation, shape, and total-density gradients}",
      journal = {\mnras},
         year = 2023,
        month = nov,
       volume = {526},
       number = {1},
        pages = {1022-1045},
          doi = {10.1093/mnras/stad2732},
archivePrefix = {arXiv},
       eprint = {2304.11712},
 primaryClass = {astro-ph.GA},
       adsurl = {https://ui.adsabs.harvard.edu/abs/2023MNRAS.526.1022L}
}

@ARTICLE{H0.ref,
       author = {{Riess}, Adam G. and {Yuan}, Wenlong and {Macri}, Lucas M. and {Scolnic}, Dan and {Brout}, Dillon and {Casertano}, Stefano and {Jones}, David O. and {Murakami}, Yukei and {Anand}, Gagandeep S. and {Breuval}, Louise and {Brink}, Thomas G. and {Filippenko}, Alexei V. and {Hoffmann}, Samantha and {Jha}, Saurabh W. and {D'arcy Kenworthy}, W. and {Mackenty}, John and {Stahl}, Benjamin E. and {Zheng}, WeiKang},
        title = "{A comprehensive measurement of the local value of the Hubble constant with 1 km s$^{-1}$ Mpc$^{-1}$ uncertainty from the Hubble Space Telescope and the SH0ES Team}",
      journal = {\apjl},
         year = 2022,
        month = jul,
       volume = {934},
       number = {1},
          eid = {L7},
        pages = {L7},
          doi = {10.3847/2041-8213/ac5c5b},
archivePrefix = {arXiv},
       eprint = {2112.04510},
 primaryClass = {astro-ph.CO},
       adsurl = {https://ui.adsabs.harvard.edu/abs/2022ApJ...934L...7R}
}

@ARTICLE{virgo.ref,
       author = {{Mould}, Jeremy R. and {Huchra}, John P. and {Freedman}, Wendy L. and {Kennicutt}, Robert C., Jr. and {Ferrarese}, Laura and {Ford}, Holland C. and {Gibson}, Brad K. and {Graham}, John A. and {Hughes}, Shaun M.~G. and {Illingworth}, Garth D. and {Kelson}, Daniel D. and {Macri}, Lucas M. and {Madore}, Barry F. and {Sakai}, Shoko and {Sebo}, Kim M. and {Silbermann}, Nancy A. and {Stetson}, Peter B.},
        title = "{The Hubble Space Telescope Key Project on the Extragalactic Distance Scale. XXVIII. Combining the constraints on the Hubble constant}",
      journal = {\apj},
         year = 2000,
        month = feb,
       volume = {529},
       number = {2},
        pages = {786-794},
          doi = {10.1086/308304},
archivePrefix = {arXiv},
       eprint = {astro-ph/9909260},
 primaryClass = {astro-ph},
       adsurl = {https://ui.adsabs.harvard.edu/abs/2000ApJ...529..786M}
}

@ARTICLE{redshift.ref,
       author = {{Falco}, Emilio E. and {Kurtz}, Michael J. and {Geller}, Margaret J. and {Huchra}, John P. and {Peters}, James and {Berlind}, Perry and {Mink}, Douglas J. and {Tokarz}, Susan P. and {Elwell}, Barbara},
        title = "{The Updated Zwicky Catalog (UZC)}",
      journal = {\pasp},
         year = 1999,
        month = apr,
       volume = {111},
       number = {758},
        pages = {438-452},
          doi = {10.1086/316343},
archivePrefix = {arXiv},
       eprint = {astro-ph/9904265},
 primaryClass = {astro-ph},
       adsurl = {https://ui.adsabs.harvard.edu/abs/1999PASP..111..438F}
}

@ARTICLE{avlaw.ref,
       author = {{Cardelli}, Jason A. and {Clayton}, Geoffrey C. and {Mathis}, John S.},
        title = "{The relationship between infrared, optical, and ultraviolet extinction}",
      journal = {\apj},
         year = 1989,
        month = oct,
       volume = {345},
        pages = {245},
          doi = {10.1086/167900},
       adsurl = {https://ui.adsabs.harvard.edu/abs/1989ApJ...345..245C}
}

@ARTICLE{Smartt2009,
       author = {{Smartt}, Stephen J.},
        title = "{Progenitors of core-collapse supernovae}",
      journal = {\araa},
         year = 2009,
        month = sep,
       volume = {47},
       number = {1},
        pages = {63-106},
          doi = {10.1146/annurev-astro-082708-101737},
archivePrefix = {arXiv},
       eprint = {0908.0700},
 primaryClass = {astro-ph.SR},
       adsurl = {https://ui.adsabs.harvard.edu/abs/2009ARA&A..47...63S}
}

@ARTICLE{Niu2023,
       author = {{Niu}, Zexi and {Sun}, Ning-Chen and {Maund}, Justyn R. and {Zhang}, Yu and {Zhao}, Ruining and {Liu}, Jifeng},
        title = "{The dusty red supergiant progenitor and the local environment of the Type II SN 2023ixf in M101}",
      journal = {\apjl},
         year = 2023,
        month = sep,
       volume = {955},
       number = {1},
          eid = {L15},
        pages = {L15},
          doi = {10.3847/2041-8213/acf4e3},
archivePrefix = {arXiv},
       eprint = {2308.04677},
 primaryClass = {astro-ph.SR},
       adsurl = {https://ui.adsabs.harvard.edu/abs/2023ApJ...955L..15N}
}

@ARTICLE{Li2011,
       author = {{Li}, Weidong and {Leaman}, Jesse and {Chornock}, Ryan and {Filippenko}, Alexei V. and {Poznanski}, Dovi and {Ganeshalingam}, Mohan and {Wang}, Xiaofeng and {Modjaz}, Maryam and {Jha}, Saurabh and {Foley}, Ryan J. and {Smith}, Nathan},
        title = "{Nearby supernova rates from the Lick Observatory Supernova Search - II. The observed luminosity functions and fractions of supernovae in a complete sample}",
      journal = {\mnras},
         year = 2011,
        month = apr,
       volume = {412},
       number = {3},
        pages = {1441-1472},
          doi = {10.1111/j.1365-2966.2011.18160.x},
archivePrefix = {arXiv},
       eprint = {1006.4612},
 primaryClass = {astro-ph.SR},
       adsurl = {https://ui.adsabs.harvard.edu/abs/2011MNRAS.412.1441L}
}

@ARTICLE{Sana2012,
       author = {{Sana}, H. and {de Mink}, S.~E. and {de Koter}, A. and {Langer}, N. and {Evans}, C.~J. and {Gieles}, M. and {Gosset}, E. and {Izzard}, R.~G. and {Le Bouquin}, J. -B. and {Schneider}, F.~R.~N.},
        title = "{Binary interaction dominates the evolution of massive stars}",
      journal = {Science},
         year = 2012,
        month = jul,
       volume = {337},
       number = {6093},
        pages = {444},
          doi = {10.1126/science.1223344},
archivePrefix = {arXiv},
       eprint = {1207.6397},
 primaryClass = {astro-ph.SR},
       adsurl = {https://ui.adsabs.harvard.edu/abs/2012Sci...337..444S}
}

@ARTICLE{Zapartas2017,
       author = {{Zapartas}, E. and {de Mink}, S.~E. and {Izzard}, R.~G. and {Yoon}, S. -C. and {Badenes}, C. and {G{\"o}tberg}, Y. and {de Koter}, A. and {Neijssel}, C.~J. and {Renzo}, M. and {Schootemeijer}, A. and {Shrotriya}, T.~S.},
        title = "{Delay-time distribution of core-collapse supernovae with late events resulting from binary interaction}",
      journal = {\aap},
         year = 2017,
        month = may,
       volume = {601},
          eid = {A29},
        pages = {A29},
          doi = {10.1051/0004-6361/201629685},
archivePrefix = {arXiv},
       eprint = {1701.07032},
 primaryClass = {astro-ph.HE},
       adsurl = {https://ui.adsabs.harvard.edu/abs/2017A&A...601A..29Z}
}

@ARTICLE{Zapartas2019,
       author = {{Zapartas}, Emmanouil and {de Mink}, Selma E. and {Justham}, Stephen and {Smith}, Nathan and {de Koter}, Alex and {Renzo}, Mathieu and {Arcavi}, Iair and {Farmer}, Rob and {G{\"o}tberg}, Ylva and {Toonen}, Silvia},
        title = "{The diverse lives of progenitors of hydrogen-rich core-collapse supernovae: the role of binary interaction}",
      journal = {\aap},
         year = 2019,
        month = nov,
       volume = {631},
          eid = {A5},
        pages = {A5},
          doi = {10.1051/0004-6361/201935854},
archivePrefix = {arXiv},
       eprint = {1907.06687},
 primaryClass = {astro-ph.HE},
       adsurl = {https://ui.adsabs.harvard.edu/abs/2019A&A...631A...5Z}
}

@misc{2025Ercolino,
  author  = {Ercolino, A and Jin, H and Langer, N and et al.},
  title   = "{The demographics of core-collapse supernovae I. The role of binary evolution and CSM interaction}",
  howpublished = {arXiv: 2510.04872},
  year    = {2025}
}

@ARTICLE{2024Dessart,
       author = {{Dessart}, Luc and {Guti{\'e}rrez}, Claudia P. and {Ercolino}, Andrea and {Jin}, Harim and {Langer}, Norbert},
        title = "{A sequence of Type Ib, IIb, II-L, and II-P supernovae from binary-star progenitors with varying initial separations}",
      journal = {\aap},
         year = 2024,
        month = may,
       volume = {685},
          eid = {A169},
        pages = {A169},
          doi = {10.1051/0004-6361/202349066},
archivePrefix = {arXiv},
       eprint = {2402.12977},
 primaryClass = {astro-ph.SR},
       adsurl = {https://ui.adsabs.harvard.edu/abs/2024A&A...685A.169D}
}

@ARTICLE{2019Renzo,
       author = {{Renzo}, M. and {Zapartas}, E. and {de Mink}, S.~E. and {G{\"o}tberg}, Y. and {Justham}, S. and {Farmer}, R.~J. and {Izzard}, R.~G. and {Toonen}, S. and {Sana}, H.},
        title = "{Massive runaway and walkaway stars. A study of the kinematical imprints of the physical processes governing the evolution and explosion of their binary progenitors}",
      journal = {\aap},
         year = 2019,
        month = apr,
       volume = {624},
          eid = {A66},
        pages = {A66},
          doi = {10.1051/0004-6361/201833297},
archivePrefix = {arXiv},
       eprint = {1804.09164},
 primaryClass = {astro-ph.SR},
       adsurl = {https://ui.adsabs.harvard.edu/abs/2019A&A...624A..66R}
}

@ARTICLE{Zapartas2021,
       author = {{Zapartas}, E. and {de Mink}, S.~E. and {Justham}, S. and {Smith}, N. and {Renzo}, M. and {de Koter}, A.},
        title = "{Effect of binary evolution on the inferred initial and final core masses of hydrogen-rich, Type II supernova progenitors}",
      journal = {\aap},
         year = 2021,
        month = jan,
       volume = {645},
          eid = {A6},
        pages = {A6},
          doi = {10.1051/0004-6361/202037744},
archivePrefix = {arXiv},
       eprint = {2002.07230},
 primaryClass = {astro-ph.HE},
       adsurl = {https://ui.adsabs.harvard.edu/abs/2021A&A...645A...6Z}
}

@ARTICLE{Hiramatsu2021,
       author = {{Hiramatsu}, Daichi and {Howell}, D. Andrew and {Moriya}, Takashi J. and {Goldberg}, Jared A. and {Hosseinzadeh}, Griffin and {Arcavi}, Iair and {Anderson}, Joseph P. and {Guti{\'e}rrez}, Claudia P. and {Burke}, Jamison and {McCully}, Curtis and {Valenti}, Stefano and {Galbany}, Llu{\'\i}s and {Fang}, Qiliang and {Maeda}, Keiichi and {Folatelli}, Gast{\'o}n and {Hsiao}, Eric Y. and {Morrell}, Nidia I. and {Phillips}, Mark M. and {Stritzinger}, Maximilian D. and {Suntzeff}, Nicholas B. and {Gromadzki}, Mariusz and {Maguire}, Kate and {M{\"u}ller-Bravo}, Tom{\'a}s E. and {Young}, David R.},
        title = "{Luminous Type II short-plateau supernovae 2006Y, 2006ai, and 2016egz: a transitional class from stripped massive red supergiants}",
      journal = {\apj},
         year = 2021,
        month = may,
       volume = {913},
       number = {1},
          eid = {55},
        pages = {55},
          doi = {10.3847/1538-4357/abf6d6},
archivePrefix = {arXiv},
       eprint = {2010.15566},
 primaryClass = {astro-ph.HE},
       adsurl = {https://ui.adsabs.harvard.edu/abs/2021ApJ...913...55H}
}

@ARTICLE{Martinez2022a,
       author = {{Martinez}, L. and {Bersten}, M.~C. and {Anderson}, J.~P. and {Hamuy}, M. and {Gonz{\'a}lez-Gait{\'a}n}, S. and {Stritzinger}, M. and {Phillips}, M.~M. and {Guti{\'e}rrez}, C.~P. and {Burns}, C. and {Contreras}, C. and {de Jaeger}, T. and {Ertini}, K. and {Folatelli}, G. and {F{\"o}rster}, F. and {Galbany}, L. and {Hoeflich}, P. and {Hsiao}, E.~Y. and {Morrell}, N. and {Orellana}, M. and {Pessi}, P.~J. and {Suntzeff}, N.~B.},
        title = "{Type II supernovae from the Carnegie Supernova Project-I. I. bolometric light curves of 74 SNe II using uBgVriYJH photometry}",
      journal = {\aap},
         year = 2022,
        month = apr,
       volume = {660},
          eid = {A40},
        pages = {A40},
          doi = {10.1051/0004-6361/202142075},
archivePrefix = {arXiv},
       eprint = {2111.06519},
 primaryClass = {astro-ph.SR},
       adsurl = {https://ui.adsabs.harvard.edu/abs/2022A&A...660A..40M}
}

@ARTICLE{Martinez2022b,
       author = {{Martinez}, L. and {Bersten}, M.~C. and {Anderson}, J.~P. and {Hamuy}, M. and {Gonz{\'a}lez-Gait{\'a}n}, S. and {F{\"o}rster}, F. and {Orellana}, M. and {Stritzinger}, M. and {Phillips}, M.~M. and {Guti{\'e}rrez}, C.~P. and {Burns}, C. and {Contreras}, C. and {de Jaeger}, T. and {Ertini}, K. and {Folatelli}, G. and {Galbany}, L. and {Hoeflich}, P. and {Hsiao}, E.~Y. and {Morrell}, N. and {Pessi}, P.~J. and {Suntzeff}, N.~B.},
        title = "{Type II supernovae from the Carnegie Supernova Project-I. II. Physical parameter distributions from hydrodynamical modelling}",
      journal = {\aap},
         year = 2022,
        month = apr,
       volume = {660},
          eid = {A41},
        pages = {A41},
          doi = {10.1051/0004-6361/202142076},
archivePrefix = {arXiv},
       eprint = {2111.06529},
 primaryClass = {astro-ph.SR},
       adsurl = {https://ui.adsabs.harvard.edu/abs/2022A&A...660A..41M}
}

@ARTICLE{2014Anderson,
       author = {{Anderson}, Joseph P. and {Gonz{\'a}lez-Gait{\'a}n}, Santiago and {Hamuy}, Mario and {Guti{\'e}rrez}, Claudia P. and {Stritzinger}, Maximilian D. and {Olivares E.}, Felipe and {Phillips}, Mark M. and {Schulze}, Steve and {Antezana}, Roberto and {Bolt}, Luis and {Campillay}, Abdo and {Castell{\'o}n}, Sergio and {Contreras}, Carlos and {de Jaeger}, Thomas and {Folatelli}, Gast{\'o}n and {F{\"o}rster}, Francisco and {Freedman}, Wendy L. and {Gonz{\'a}lez}, Luis and {Hsiao}, Eric and {Krzemi{\'n}ski}, Wojtek and {Krisciunas}, Kevin and {Maza}, Jos{\'e} and {McCarthy}, Patrick and {Morrell}, Nidia I. and {Persson}, Sven E. and {Roth}, Miguel and {Salgado}, Francisco and {Suntzeff}, Nicholas B. and {Thomas-Osip}, Joanna},
        title = "{Characterizing the V-band light-curves of hydrogen-rich Type II supernovae}",
      journal = {\apj},
         year = 2014,
        month = may,
       volume = {786},
       number = {1},
          eid = {67},
        pages = {67},
          doi = {10.1088/0004-637X/786/1/67},
archivePrefix = {arXiv},
       eprint = {1403.7091},
 primaryClass = {astro-ph.HE},
       adsurl = {https://ui.adsabs.harvard.edu/abs/2014ApJ...786...67A}
}

@ARTICLE{Martinez2022c,
       author = {{Martinez}, L. and {Anderson}, J.~P. and {Bersten}, M.~C. and {Hamuy}, M. and {Gonz{\'a}lez-Gait{\'a}n}, S. and {Orellana}, M. and {Stritzinger}, M. and {Phillips}, M.~M. and {Guti{\'e}rrez}, C.~P. and {Burns}, C. and {de Jaeger}, T. and {Ertini}, K. and {Folatelli}, G. and {F{\"o}rster}, F. and {Galbany}, L. and {Hoeflich}, P. and {Hsiao}, E.~Y. and {Morrell}, N. and {Pessi}, P.~J. and {Suntzeff}, N.~B.},
        title = "{Type II supernovae from the Carnegie Supernova Project-I. III. Understanding SN II diversity through correlations between physical and observed properties}",
      journal = {\aap},
         year = 2022,
        month = apr,
       volume = {660},
          eid = {A42},
        pages = {A42},
          doi = {10.1051/0004-6361/202142555},
archivePrefix = {arXiv},
       eprint = {2202.11220},
 primaryClass = {astro-ph.SR},
       adsurl = {https://ui.adsabs.harvard.edu/abs/2022A&A...660A..42M}
}

@ARTICLE{Sun2021,
       author = {{Sun}, Ning-Chen and {Maund}, Justyn R. and {Crowther}, Paul A. and {Fang}, Xuan and {Zapartas}, Emmanouil},
        title = "{Towards a better understanding of supernova environments: a study of SNe 2004dg and 2012P in NGC 5806 with HST and MUSE}",
      journal = {\mnras},
         year = 2021,
        month = jun,
       volume = {504},
       number = {2},
        pages = {2253-2272},
          doi = {10.1093/mnras/stab994},
archivePrefix = {arXiv},
       eprint = {2011.13667},
 primaryClass = {astro-ph.SR},
       adsurl = {https://ui.adsabs.harvard.edu/abs/2021MNRAS.504.2253S}
}

@ARTICLE{Teja2023,
       author = {{Teja}, Rishabh Singh and {Singh}, Avinash and {Sahu}, D.~K. and {Anupama}, G.~C. and {Kumar}, Brajesh and {Nakaoka}, Tatsuya and {Kawabata}, Koji S. and {Yamanaka}, Masayuki and {Takey}, Ali and {Kawabata}, Miho},
        title = "{SN 2018gj: a short plateau type II supernova with persistent blueshifted Ha emission}",
      journal = {\apj},
         year = 2023,
        month = sep,
       volume = {954},
       number = {2},
          eid = {155},
        pages = {155},
          doi = {10.3847/1538-4357/acdf5e},
archivePrefix = {arXiv},
       eprint = {2306.10136},
 primaryClass = {astro-ph.HE},
       adsurl = {https://ui.adsabs.harvard.edu/abs/2023ApJ...954..155T}
}

@ARTICLE{Farrell2020,
       author = {{Farrell}, Eoin J. and {Groh}, Jose H. and {Meynet}, Georges and {Eldridge}, J.~J.},
        title = "{The uncertain masses of progenitors of core-collapse supernovae and direct-collapse black holes}",
      journal = {\mnras},
         year = 2020,
        month = may,
       volume = {494},
       number = {1},
        pages = {L53-L58},
          doi = {10.1093/mnrasl/slaa035},
archivePrefix = {arXiv},
       eprint = {2001.08711},
 primaryClass = {astro-ph.SR},
       adsurl = {https://ui.adsabs.harvard.edu/abs/2020MNRAS.494L..53F}
}

@article{Izzard+2004,
   author = {R.~G. Izzard and C.~A. Tout and A.~I. Karakas and O.~R. Pols},
   doi = {10.1111/j.1365-2966.2004.07446.x},
   journal = {\mnras},
   month = {5},
   pages = {407-426},
   title = "{A new synthetic model for asymptotic giant branch stars}",
   volume = {350},
   year = {2004},
}

@ARTICLE{Zapartas2017b,
       author = {{Zapartas}, E. and {de Mink}, S.~E. and {Van Dyk}, S.~D. and {Fox}, O.~D. and {Smith}, N. and {Bostroem}, K.~A. and {de Koter}, A. and {Filippenko}, A.~V. and {Izzard}, R.~G. and {Kelly}, P.~L. and {Neijssel}, C.~J. and {Renzo}, M. and {Ryder}, S.},
        title = "{Predicting the presence of companions for stripped-envelope supernovae: the case of the Broad-lined Type Ic SN 2002ap}",
      journal = {\apj},
         year = 2017,
        month = jun,
       volume = {842},
       number = {2},
          eid = {125},
        pages = {125},
          doi = {10.3847/1538-4357/aa7467},
archivePrefix = {arXiv},
       eprint = {1705.07898},
 primaryClass = {astro-ph.HE},
       adsurl = {https://ui.adsabs.harvard.edu/abs/2017ApJ...842..125Z}
}

@article{Izzard+2006,
   author = {R.~G. Izzard and L.~M. Dray and A.~I. Karakas and M Lugaro and C.~A. Tout},
   doi = {10.1051/0004-6361:20066129},
   journal = {\aap},
   month = {12},
   pages = {565-572},
   title = "{Population nucleosynthesis in single and binary stars. I. model}",
   volume = {460},
   year = {2006},
}

@article{Izzard+2009,
   author = {R.~G. Izzard and E Glebbeek and R.~J. Stancliffe and O.~R. Pols},
   doi = {10.1051/0004-6361/200912827},
   journal = {\aap},
   month = {12},
   pages = {1359-1374},
   title = "{Population synthesis of binary carbon-enhanced metal-poor stars}",
   volume = {508},
   year = {2009},
}

@ARTICLE{Auchettl+2018,
       author = {{Auchettl}, Katie and {Lopez}, Laura A. and {Badenes}, Carles and {Ramirez-Ruiz}, Enrico and {Beacom}, John F. and {Holland-Ashford}, Tyler},
        title = "{Measurement of the Core-collapse progenitor mass distribution of the Small Magellanic Cloud}",
      journal = {\apj},
         year = 2019,
        month = jan,
       volume = {871},
       number = {1},
          eid = {64},
        pages = {64},
          doi = {10.3847/1538-4357/aaf395},
archivePrefix = {arXiv},
       eprint = {1804.10210},
 primaryClass = {astro-ph.SR},
       adsurl = {https://ui.adsabs.harvard.edu/abs/2019ApJ...871...64A}
}

@ARTICLE{Murphy+2024,
       author = {{Murphy}, Jeremiah W. and {Barrientos}, Andr{\'e}s F. and {Andrae}, Ren{\'e} and {Guzman}, Joseph and {Williams}, Benjamin F. and {Dalcanton}, Julianne J. and {Koplitz}, Brad},
        title = "{The mass of the Vela Pulsar progenitor and the age of the Vela-Puppis complex}",
      journal = {\apj},
         year = 2025,
        month = aug,
       volume = {988},
       number = {2},
          eid = {241},
        pages = {241},
          doi = {10.3847/1538-4357/ade5b4},
archivePrefix = {arXiv},
       eprint = {2406.04075},
 primaryClass = {astro-ph.SR},
       adsurl = {https://ui.adsabs.harvard.edu/abs/2025ApJ...988..241M}
}

@ARTICLE{Schootemeijer+2018,
       author = {{Schootemeijer}, A. and {G{\"o}tberg}, Y. and {de Mink}, S.~E. and {Gies}, D. and {Zapartas}, E.},
        title = "{Clues about the scarcity of stripped-envelope stars from the evolutionary state of the sdO+Be binary system {\ensuremath{\varphi}} Persei}",
      journal = {\aap},
         year = 2018,
        month = jul,
       volume = {615},
          eid = {A30},
        pages = {A30},
          doi = {10.1051/0004-6361/201731194},
archivePrefix = {arXiv},
       eprint = {1803.02379},
 primaryClass = {astro-ph.SR},
       adsurl = {https://ui.adsabs.harvard.edu/abs/2018A&A...615A..30S}
}

@ARTICLE{2017gotberg,
       author = {{G{\"o}tberg}, Y. and {de Mink}, S.~E. and {Groh}, J.~H.},
        title = "{Ionizing spectra of stars that lose their envelope through interaction with a binary companion: role of metallicity}",
      journal = {\aap},
         year = 2017,
        month = nov,
       volume = {608},
          eid = {A11},
        pages = {A11},
          doi = {10.1051/0004-6361/201730472},
archivePrefix = {arXiv},
       eprint = {1701.07439},
 primaryClass = {astro-ph.SR},
       adsurl = {https://ui.adsabs.harvard.edu/abs/2017A&A...608A..11G}
}

@ARTICLE{2024Hovis,
       author = {{Hovis-Afflerbach}, B. and {G{\"o}tberg}, Y. and {Schootemeijer}, A. and {Klencki}, J. and {Strom}, A.~L. and {Ludwig}, B.~A. and {Drout}, M.~R.},
        title = "{The mass distribution of stars stripped in binaries: the effect of metallicity}",
      journal = {arXiv e-prints},
         year = 2024,
        month = dec,
          eid = {arXiv:2412.05356},
        pages = {arXiv:2412.05356},
          doi = {10.48550/arXiv.2412.05356},
archivePrefix = {arXiv},
       eprint = {2412.05356},
 primaryClass = {astro-ph.SR},
       adsurl = {https://ui.adsabs.harvard.edu/abs/2024arXiv241205356H}
}

@ARTICLE{Castrillo+2020,
       author = {{Castrillo}, Asier and {Ascasibar}, Yago and {Galbany}, Llu{\'\i}s and {S{\'a}nchez}, Sebasti{\'a}n F. and {Badenes}, Carles and {Anderson}, Joseph P. and {Kuncarayakti}, Hanindyo and {Lyman}, Joseph D. and {D{\'\i}az}, Angeles I.},
        title = "{The delay time distribution of supernovae from integral-field spectroscopy of nearby galaxies}",
      journal = {\mnras},
         year = 2021,
        month = mar,
       volume = {501},
       number = {3},
        pages = {3122-3136},
          doi = {10.1093/mnras/staa3876},
archivePrefix = {arXiv},
       eprint = {2012.11958},
 primaryClass = {astro-ph.GA},
       adsurl = {https://ui.adsabs.harvard.edu/abs/2021MNRAS.501.3122C}
}

@article{Vink+2001,
   author = {Jorick S. Vink and A. de Koter and H. J. G. L. M. Lamers},
   doi = {10.1051/0004-6361:20010127},
   issn = {0004-6361},
   issue = {2},
   journal = {Astronomy and Astrophysics},
   month = {4},
   pages = {574-588},
   title = "{Mass-loss predictions for O and B stars as a function of metallicity}",
   volume = {369},
   url = {http://adsabs.harvard.edu/abs/2001A%26A...369..574V},
   year = {2001},
}

@ARTICLE{2017Yoon,
       author = {{Yoon}, Sung-Chul and {Dessart}, Luc and {Clocchiatti}, Alejandro},
        title = "{Type Ib and IIb supernova progenitors in interacting binary systems}",
      journal = {\apj},
         year = 2017,
        month = may,
       volume = {840},
       number = {1},
          eid = {10},
        pages = {10},
          doi = {10.3847/1538-4357/aa6afe},
archivePrefix = {arXiv},
       eprint = {1701.02089},
 primaryClass = {astro-ph.SR},
       adsurl = {https://ui.adsabs.harvard.edu/abs/2017ApJ...840...10Y}
}

@book{Ivanova+2020,
  author    = {Ivanova, Natalia and Justham, Stephen and Ricker, Paul},
  title     = "{Common Envelope Evolution}",
  publisher = {IOP Publishing},
  address   = {Bristol, UK}, 
  year      = {2020},
  url       = {https://doi.org/10.1088/2514-3433/abb6f0},
  doi       = {10.1088/2514-3433/abb6f0}
}

@ARTICLE{Ertl+2016,
       author = {{Ertl}, T. and {Janka}, H. -Th. and {Woosley}, S.~E. and {Sukhbold}, T. and {Ugliano}, M.},
        title = "{A two-parameter criterion for classifying the explodability of massive stars by the neutrino-driven mechanism}",
      journal = {\apj},
         year = 2016,
        month = feb,
       volume = {818},
       number = {2},
          eid = {124},
        pages = {124},
          doi = {10.3847/0004-637X/818/2/124},
archivePrefix = {arXiv},
       eprint = {1503.07522},
 primaryClass = {astro-ph.SR},
       adsurl = {https://ui.adsabs.harvard.edu/abs/2016ApJ...818..124E}
}

@ARTICLE{2007Crowther,
       author = {{Crowther}, Paul A.},
        title = "{Physical properties of Wolf-Rayet stars}",
      journal = {\araa},
         year = 2007,
        month = sep,
       volume = {45},
       number = {1},
        pages = {177-219},
          doi = {10.1146/annurev.astro.45.051806.110615},
archivePrefix = {arXiv},
       eprint = {astro-ph/0610356},
 primaryClass = {astro-ph},
       adsurl = {https://ui.adsabs.harvard.edu/abs/2007ARA&A..45..177C}
}

\end{document}